\documentclass[USenglish]{article}	
\usepackage[utf8]{inputenc}				%(only for the pdftex engine)
\usepackage[big,online]{dgruyter}	%values: small,big | online,print,work
\usepackage{lmodern} 
\usepackage{microtype}
\usepackage[numbers,square,sort&compress]{natbib}

\usepackage{xcolor}
\usepackage{graphicx}
\usepackage{amsmath}
\usepackage{amssymb}
\usepackage{caption}
\usepackage{subcaption}
\usepackage{float}
\usepackage{booktabs}

\theoremstyle{dgthm}

\theoremstyle{dgdef}

\begin{document}

\newcommand{\y}{\boldsymbol{y}}
\newcommand{\ty}{\tilde{\y}}
\newcommand{\blambda}{\boldsymbol{\lambda}}
\newcommand{\btau}{\boldsymbol{\tau}}
\newcommand{\btheta}{\boldsymbol{\theta}}
\newcommand{\lat}{\text{lat}}
\newcommand{\f}{\text{for}}
\newcommand{\bx}{\boldsymbol{X}}
\newcommand{\yf}{\ty^{\f}}
\newcommand{\yl}{\ty^{\lat}}
\newcommand{\tl}{\btheta^{\lat}}
\newcommand{\tf}{\btheta^{\f}}
\newcommand{\Xl}{\bx^{\lat}}
\newcommand{\Xf}{\bx^{\f}}
\newcommand{\I}{\boldsymbol{I}}

	%\include{custom_commands.tex}
%%%--------------------------------------------%%%
	\articletype{Research Article}
	\received{Month	DD, YYYY}
	\revised{Month	DD, YYYY}
  \accepted{Month	DD, YYYY}
  \journalname{JQAS}
  \journalyear{YYYY}
  \journalvolume{XX}
  \journalissue{X}
  \startpage{1}
  \aop
  \DOI{10.1515/sample-YYYY-XXXX}
%%%--------------------------------------------%%%

\title{A generative approach to frame-level multi-competitor races}
\runningtitle{Frame-level Multi-competitor models}
%\subtitle{Insert subtitle if needed}

\author*[1,4]{Tyrel Stokes}
%\ use * to mark the author as the corresponding author
\author[2]{Gurashish Bagga}
\author[2]{Kimberly Kroetch} 
\author[2,4]{Brendan Kumagai} 
\author[3]{Liam Welsh} 
\runningauthor{Stokes et al.}
\affil[1]{\protect\raggedright 
NYU Langone, Department of Biostatistics, New York, USA, e-mail: tyrel.stokes@nyulangone.org}
\affil[2]{\protect\raggedright 
Simon Fraser University, Department of Statistics and Actuarial Science, Burnaby, Canada, e-mail: gurashish\_bagga@sfu.ca, kimberly\_kroetch@sfu.ca, brendan\_kumagai@sfu.ca}
\affil[3]{\protect\raggedright 
University of Toronto, Department of Statistical Sciences, Toronto, Canada, e-mail: liam.welsh@mail.utoronto.ca}
\affil[4]{\protect\raggedright 
Zelus Analytics}
	
%\communicated{...}
%\dedication{...}
	
\abstract{Multi-competitor races often feature complicated within-race strategies that are difficult to capture when training data on race outcome level data. Models which do not account for race-level strategy may suffer from confounded inferences and predictions. We develop a generative model for multi-competitor races which explicitly models race-level effects like drafting and separates strategy from competitor ability. The model allows one to simulate full races from any real or created starting position opening new avenues for attributing value to within-race actions and performing counter-factual analyses. This methodology is sufficiently general to apply to any track based multi-competitor races where both tracking data is available and competitor movement is well described by simultaneous forward and lateral movements. We apply this methodology to one-mile horse races using frame-level tracking data provided by the New York Racing Association (NYRA) and the New York Thoroughbred Horsemen's Association (NYTHA) for the Big Data Derby 2022 Kaggle Competition.  We demonstrate how this model can yield new inferences, such as the estimation of horse-specific speed profiles and examples of posterior predictive counterfactual simulations to answer questions of interest such as starting lane impacts on race outcomes.}


    \maketitle

\section{Introduction} 

In multi-competitor sports, athletes and teams want not only to understand the relative performance and underlying abilities of competitors, but also better understand optimal within-race strategies to help a competitor improve. In highly strategic races, such as middle distance running or our canonical example of horse racing, teasing apart within-race strategic effects from the underlying abilities of competitors is extremely difficult using current methods which are trained using race-level outcomes. Optimal strategy in such races is likely to depend not only on the quality of the competitors, but also on particularities of each race including the weather conditions and even within-race conditions such as particular competitors getting good starts or a competitor having restricted movement due to surrounding competitors. Further, traditional analyses which operate on race-level statistics like finishing time may easily be confounded with respect to estimating competitor ability since competitors of similar quality may be more likely to race against each other and the optimal strategies for each competitor given their competitors are likely to vary according to their own abilities. For example, an elite NCAA middle distance runner might typically prefer a front running strategy where they attempt to lead the race with a fast enough pace to drop their opponents, but they might not be fast enough for this strategy to be optimal in a semi-final or final of the world championships. Without methods capable of teasing strategy and ability apart, counterfactual analysis aiming to estimate what might have occurred under different strategies or inferring underlying ability are  likely to be unreliable. This leaves coaches, athletes, and teams in an information deficit with respect to where they stand relative to their competitors and what they might be able to achieve.\\

In this paper, we extend recent work in modelling continuous outcomes in multi-competitor games \citep{che2022athlete} to the context of frame-level tracking data. In our canonical example of horse racing, we capture the interdependent strategic effects of the competitors by simultaneously modelling forward and horizontal movement as a function of underlying ability and relative spatial positioning with respect to all other competitors. We propose a generative Bayesian model which allows one to take advantage of posterior predictive simulation. In particular, this allows one to simulate counter-factual races and scenarios. For example, one can simulate races with competitors who have not necessarily raced against each other. The framework is rich enough to simulate alternative strategies by one or more competitors, estimate their impact on performance, and estimate the impact of race conditions outside of the competitor's control such as the impact of starting lanes on finishing probabilities.

\section{Extending Dynamic linear Models to multi-competitor Frame-Level Competitions}\label{sc:method}

Much of the literature in multi-competitor sports has focused on modelling rank-type data \citep{plackett1975analysis,harville1973assigning,henery1981permutation,luce1959individual}  and more recently
\cite{glickman2015stochastic} incorporating these ranking models into a dynamic state-space framework where latent competitor abilities evolve over time. This dynamic state-space approach to allowing competitor abilities to evolve over time was originally developed in the context of head-to-head games or paired comparisons \citep{fahrmeir1994dynamic,glickman1999parameter,glickman2001dynamic,glickman2005state}. One of the advantages of working directly with ranks as opposed to other continuous measures of success or performance, besides the ubiquity of this kind of data across a multitude of competitions, is they may be more robust to certain strategic effects. For example a runner may choose to run sub-maximally against weaker competition, particularly in earlier rounds or heats and training a model on run times directly may produce misleading predictions as a result. On the other hand, excluding data in earlier rounds of competition or in cases where there may be incentives not aligned with producing maximal continuous outcome results may result in severely shrinking the pool of competitors over which one can learn relative abilities. The cost, however, of modelling ranks directly is coarsening the data used in the modelling step and potential loss of information. More recent work has explored adding information from continuous outcomes for head-to-head competitions \citep{kovalchik2020extension} and Glickman and Che \citep{che2022athlete} proposed an extension for multi-competitor sports. The key idea in \citep{che2022athlete} is to learn a transformation of the continuous outcome and to control for game-specific and potentially strategic effects using covariates and functions of the latent competitor abilities. In particular, they propose using dynamic linear models (DLMs) with (monotonic) transformed outcomes which are in part learned from the data. This approach allows one to balance the simplicity of the DLM framework while maintaining the flexibility necessary to model arbitrary multi-competitor sports competitions. Consider the probability model:
\begin{align}
    p(\btau_{\blambda}(\ty)|\btheta_t,\bx,\sigma),
\end{align}
where $\btau(\cdot)$ represents a (learned) transformation function of a pre-processed outcome vector, $\ty$, and $\btheta_t$ represents a vector of competitor ability parameters at time $t$, $\bx$ is a set of competition level covariates, and $\sigma$ is a noise parameter. Following previous work in competitor ratings, such as \cite{glickman1999parameter,glickman2001dynamic,glickman2005state,fahrmeir1994dynamic,glickman2015stochastic}, the competitor abilities are allowed to evolve over time using stochastic process priors such as a random walk.\\

In the context of frame-level data, we often have 1-25 frames of data per second with the locations of all competitors recorded at each frame. In this work, we are interested both in recovering competition-level predictions, such as winning times and competitor ranking, and also having a rich enough framework to simulate counter-factual scenarios and strategies. Simulating entire races and capturing the strategic nuances of multi-competitor racing requires generating predicted locations at every frame until the simulated race is over. This rules out rank-like models at the frame-level since they are unable to reproduce the locations of the competitors in each frame in a generative sense. The goal is then adapting the Glickman and Che \cite{che2022athlete} framework by modelling directly a function of competitor location at each time, taking into consideration in-game and in-frame strategic effects. The key idea in this framework to properly account for the strategic components in such races as well as specifying a model rich enough to generate exact locations along the track is to split the movement of each competitor in each frame into two components - a forward distance, $\yf$, and a lateral distance, $\yl$. We define forward distance to be distance travelled perpendicular to the inside of the track and we refer to forward and perpendicular distance interchangeably. Under this definition, a one-mile race is completed by a competitor once they have travelled exactly one-mile in terms of forward (or perpendicular) distance. Lateral distance, then, is defined to be the complimentary movement inside or outside of the track with respect to the forward or perpendicular distance. By transforming the distances to be relative to the inside of the track the lateral distance can be thought of in terms of lane. When possible competitors prefer to run closer to the inside of the track since this decreases the total distance needed to cover over the duration of the race, all else being equal.\\

\begin{figure}[!htb]
\centering
\includegraphics[width=0.7\textwidth]{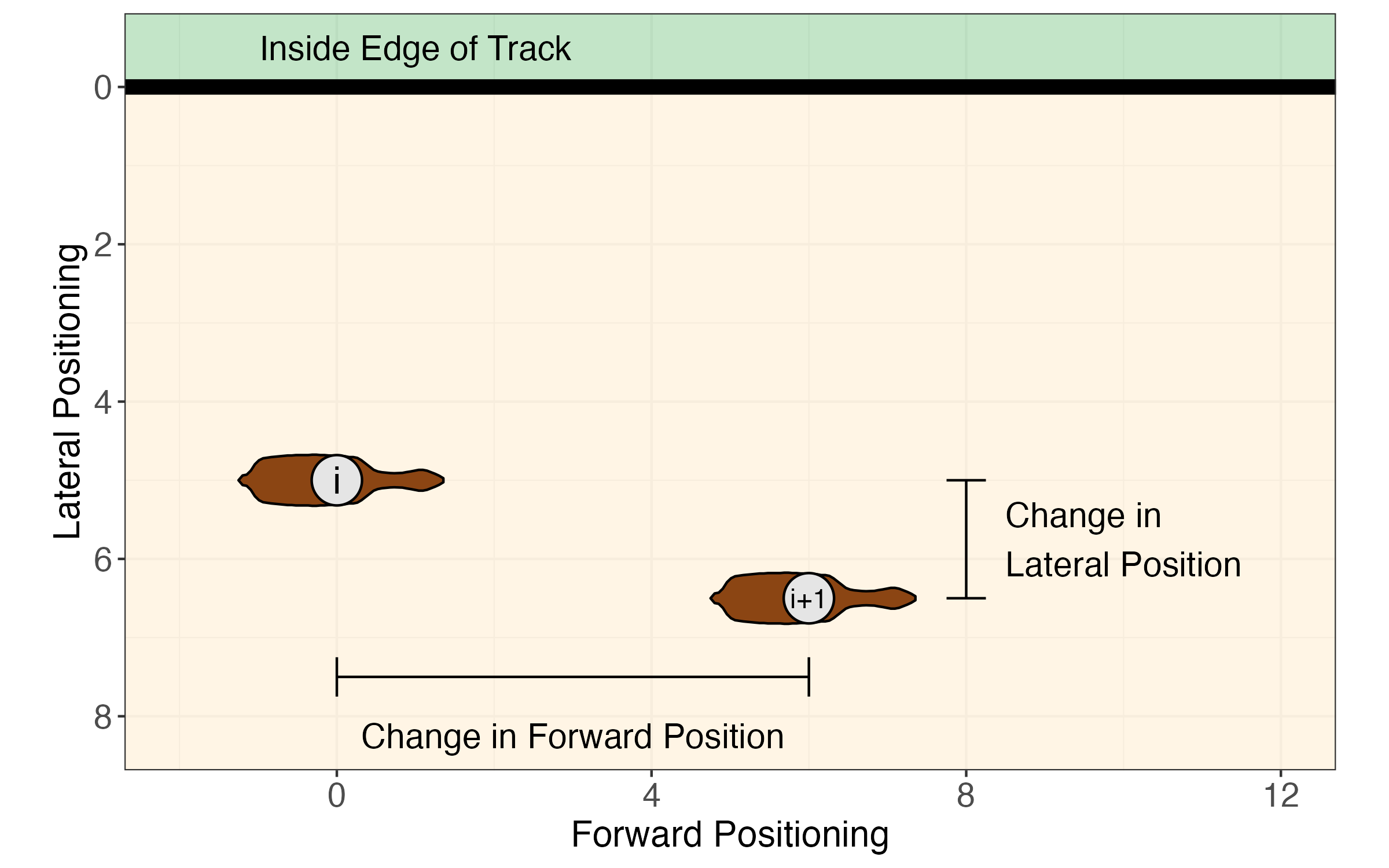}
\caption{\label{fig:fwdlatExample} An illustration of the change in a horse's forward and lateral positioning from frame labelled $i$ and the subsequent frame $i + 1$.  The lateral distance is calculated with respect to movement towards or away from the inside of track. Forward distance is thus any movement perpendicular to the inside of the track. In track based sports much of the maneuvering is with respect to guarding the lane positioning since traveling near the inside of the track allows the competitor to cover less distance over the course of the race. The transformation of relative coordinates to forward and lateral distance allows us to represent the movement in strategically relevant terms.}
\end{figure}

Figure \ref{fig:fwdlatExample} provides an illustration of the change in a horse's forward and lateral positioning between frames. In this figure, the horse moves from (0, 5) in the forward-lateral positioning plane to (5, 6.5) between frames $i$ and $i+1$. This corresponds to a forward distance travelled of 5 metres and a lateral distance travelled of 1.5 metres. The primary advantage of converting to a coordinate system which is relative to the inside track and thus lane positioning is that the movements can be consistently defined over the whole track including the turns.

Under those definitions, total distance travelled in each frame is then a simple function of the forward and lateral distance ($\tilde{\y} = \sqrt{(\yf)^2 + (\yl)^2}$). In principle each of these components can be transformed following Glickman and Che \citep{che2022athlete}, but for simplicity, in this text we will consider a simple known transformation where we model the additional distance travelled forward and laterally in each frame. The goal then is to model the following joint distribution: 
\begin{align}
p(\yf_i,\yl_i|\tf(j),\tl(j),\Xl_{i},\Xf_{i},\Sigma, \psi),  i = 1,2,\dots,\I,\label{eq:gen mod}
\end{align}
where $i$ represents an arbitrary frame which increases to the vector of random variables $\I$ which represents the final frame for each of the competitors, $\tl$ and $\tf$ are within-race competitor ability vectors, $\Xl_{i}$ and $\Xf_{i}$ are covariates, $\Sigma$ is a variance-covariance matrix, and $\psi$ is a vector of covariate coefficients. The competitor ability vectors, $\tl$ and $\tf$, depend on where in the race the competitors find themselves in the race at frame $i$. We denote the distance travelled up to frame $i$ by the index $j$. This allows us to model the competitors ability across different phases of the race including reaction to the starting gun, initial acceleration, drive and maintenance phases, as well as the final stretch for example. In many racing sports we might expect there to be different types of competitors which excel at different phases which is not well captured by an overall constant level of ability throughout the race. To model competitor ability throughout the different race phases parsimoniously we assume that the underlying ability evolves continuously and smoothly over the course of the race. Specifically, we use a spline based approach to reduce the continuous ability vectors to a (relatively small) finite dimensional set of basis parameters. One pre-specifies a number of knots or degrees of freedom and for each competitor, $k$, their forward or lateral ability is represented by a finite vector of parameters, ($\beta_1,\dots,\beta_{d})^k$, where $d$ is the dimension of the competitor-level within-race coefficients. The continuous coefficients are smooth functions of the finite vector representation. In addition to providing more nuanced simulation possibilities, the estimated within-race competitor-specific coefficients allows one to characterize notions of both ability and style. In Section \ref{sc: profile} we discuss how one can cluster within-race coefficients in the context of horse racing to reveal racing styles or competitor profiles.\\ 

In the previous paragraph we discussed how in our canonical horse racing example the index $j$ represents the cumulative distance that a competitor has travelled up until frame $i$. It is important to note that this is only one of potentially several ways to index a race or race phase. One could alternatively imagine using total time elapsed up until frame $i$, for example. Across different race types, different indexing may correspond better or worse to the underlying latent race phase and this indexing should be chosen with the guidance of domain experts to insure the estimated race-varying effects are meaningful.\\

Further note that although suppressed in the notation here for simplicity, these competitor ability vectors may depend on some time period $t$ and the spline vectors can be updated according to a stochastic process prior in a way similar to that which is standard in the dynamic competitor rating literature as discussed above. For computational simplicity we propose modelling the joint distribution of an appropriate transformation of the frame-level forward and lateral distances with normal  or truncated normal distributions, where the matrix $\Sigma$ represents the variance-covariance matrix. This can be easily extended to more specific distributional choices when computational resources permit. It may be especially important to consider probability models which bound lateral movement according to lane constraints of the track when simulating some race types, but for simplicity we leave this as an extension.\\

$\Xl_{i},\Xf_{i}$ represent the lateral and forward covariates, respectively. The covariates can be categorized into two groups - dynamic covariates which change over the course of the race and static race-level covariates. The most important spatial covariates capture interactions between competitors. For example, we may expect competitors far ahead of the field to slow up near the end of a race or for a racer who is boxed in to be more restricted in the type of movement they can make. In Section \ref{sc:drafting} we discuss Drafting variables. This is a particularly difficult dynamic feature in that it depends on the relative position of all competitors and there may be both short-term and long-term effects. For example, we might expect a competitor to expend additional energy to close a gap in order to more effectively draft in the short-term and in the long-term we might expect having drafted more effectively in the past may lead to more energy and speed in later stages of the race. Additionally in Section \ref{sc:mod and sim} we discuss using simple spatial representations of relative forward/backward and side-to-side positions of horses to predict lateral movement. These kinds of covariates are crucial to capture and effectively simulate strategic behaviour.\\

Once a model for equation \eqref{eq:gen mod} has been proposed and fitted, one can simulate full races. Bayesian, or approximately Bayesian, procedures naturally allow one to account for uncertainty in both the generative procedure and uncertainty with respect to unknown parameters via posterior predictive simulation and is our focus in this article. Modelling the joint distribution for all competitors' forward and lateral movement in each frame allows us to perform several new kinds of simulation analyses to better understand performance and strategies in complex multi-competitor races. Two notable types of simulation analyses are within-race value attribution and counter-factual analysis.\\ 

In continuous team sports there has been a recent emphasis on models which generate instantaneous notions of value, notably the landmark basketball paper by Cervone et al. \citep{cervone2016multiresolution} which formalized the notion of an Expected Possession Value (EPV) in basketball, which has since been adapted to other continuous sports including soccer \citep{fernandez2021framework}. The idea is to model the future actions and rewards of those actions given all of the (spatial) information present at a given moment to generate a value for the possession averaging over the possible future evolutions of the possession. This is represented mathematically as:
\begin{align}
E[\bx|\mathcal{F}_t] = \int_{\omega}\bx(\omega)P(d\omega|\mathcal{F}_t),
\end{align}
where $\bx$ is a value outcome of interest, $\omega$ is a path or possession path, and $\mathcal{F}_t$ is a sigma-algebra representing the (spatial) information up to time $t$.\\

One of the difficulties of these continuous sports is that the actions and strategies that we would like to value often take place disconnected in space and time from the subsequent rewards. This makes it especially difficult to say how valuable a particular pass was or the cost of turning over the ball, for example, might be. The EPV framework solves this problem by converting spatial information into a continuous stock-ticker of value. Actions, and changes in spatial positioning, have impacts on the future evolutions of the play which are then captured by changes in EPV and these changes, or deltas, can be attributed to competitors or strategies through actions and/or functions of spatial positioning. Generally, continuous actions sports like basketball, soccer, and hockey are too complicated to simulate at a generative level and instead approximations must be made to estimate instantaneous notions of value. We show in our horse racing example in the sections to follow that our proposed simulation framework for multi-competitor races is both rich enough to simulate entire races with uncertainty and computationally feasible. This means that like the EPV framework, we can generate instantaneous values, such as expectation over race finishing time or ranking for each competitor, but additionally we can actually reproduce an entire set of sample future paths. In mathematical terms, value outcomes of interest like finishing time will be some deterministic function, $h(\cdot)$, of the entire history of forward locations, $\yf_{1:I}$. Adapting the notation from Cervone et al. \citep{cervone2016multiresolution} to our context we can express the posterior predictive of the forward position conditional of information up to some specified frame $i$ as 
\begin{align}
p(\yf_{1:I}|\mathcal{F}_i,(\boldsymbol{Y},\boldsymbol{X})) = \int\int\int p(\yf_{1:s},\yl_{1:s},I=s|\mathcal{F}_i,\gamma)f(\gamma|(\boldsymbol{Y},\boldsymbol{X}))d\yl(s)dsd\gamma,
\end{align}
where $\gamma = (\tf(j),\tl(j),\Sigma, \psi)$ are all of the parameters in equation \eqref{eq:gen mod}, $(\boldsymbol{Y},\boldsymbol{X})$ is all of the data used to fit the posterior, $\mathcal{F}_i$ represents the information available in frame $i$ for the simulation at hand, and $I$ is the vector of final frames for all participants. We can think of $I$ as a vector of stopping times equivalent to possession stopping times in the EPV framework. Any outcome of interest, such as finishing time or rank or rank up to a certain point of the race past frame $i$ will be a deterministic function of this distribution. Collaboration with experts would allow us to design metrics and models based on the changes in finishing time and ranking to better value and understand competitor choices with regards to making outside moves or drafting and/or pinpoint where in a race a competitor lost or gained future positioning.\\

In addition to absorbing many of the benefits of the EPV framework, the relative simplicity of many multi-competitor races allows us to also perform plausible counter-factual analyses. As mentioned above, as we demonstrate in our horse racing example, it is possible to fully simulate entire multi-competitor races starting from any position. In principle, this allows one with the collaboration of experts to fix strategies of particular competitors and to average over race outcomes to value those strategies. For example, one could start near the end of a race where a competitor decides to take an outside lane to overtake a competitor. One could estimate the probability of winning for each competitor had they waited any number of meters to make their move. These kinds of analyses, at the frame-level granularity of producing not only distributions of outcomes but distributions of race paths for all competitors, is largely computationally infeasible at scale for most team-level sports due to the additional dimensionality of the competitor movement and action spaces. This is true even in those sports for which EPV has been well established, like basketball or soccer. We believe this offers a unique opportunity for better understanding multi-competitor races since it allows us to examine and represent uncertainty over value outcomes of interest and additionally allows us to directly study the properties of the produced sample paths, which may be especially useful in strategic races.

\section{Application to Horse Racing}

Horse racing is an example of a multi-competitor race with dynamic and complex intra-race strategies. For example, a jockey  may need to conserve their horse's energy via drafting while avoiding their horse getting boxed in by competitors and losing position. Given the costs and complexity of horse racing, statistical models capable of better understanding and valuing horses, jockeys, and strategies can greatly benefit owners and team members by providing insights into their horses. Through Kaggle's Big Data Derby 2022 \citep{bdd}, sponsored by the New York Racing Association (NYRA) and the New York Thoroughbred Horsemen's Association (NYTHA), we obtained tracking data recording the longitude and latitude positions of all competing horses at a frequency of approximately 4 frames/second. The data set includes all NYRA races from the 2019 season at Aqueduct Racetrack, Belmont Park, and Saratoga Race Course.\\

The goal of this application is to demonstrate the viability of the framework outlined in Section \ref{sc:method} and it's versatility for answering a variety of complicated questions not adequately addressed by methods which focus on race-level outcomes. Specifically, at the frame-level we wish to predict the future position of each horse given their current position on the track and with respect to their competitors. Doing so, we are able to develop a race simulation at any frame in the race and compute placement (e.g. 1st, 2nd, 3rd, etc.) probabilities for each horse which converge to the true result as the races progresses.

\subsection{Data Preparation}\label{sc:data prep}

We perform multiple operations in order to transform the data to suit our needs. The primary challenges we had to tackle in order to prepare our data for modelling and analyses were gathering data for track outlines and finish lines, converting coordinates from longitude and latitude to Cartesian coordinates, partitioning the track into stretches and turns, and imputing missing or incorrect data.\\

\begin{figure}
     \centering
     \begin{subfigure}[b]{0.45\textwidth}
         \centering         \includegraphics[width=\textwidth]{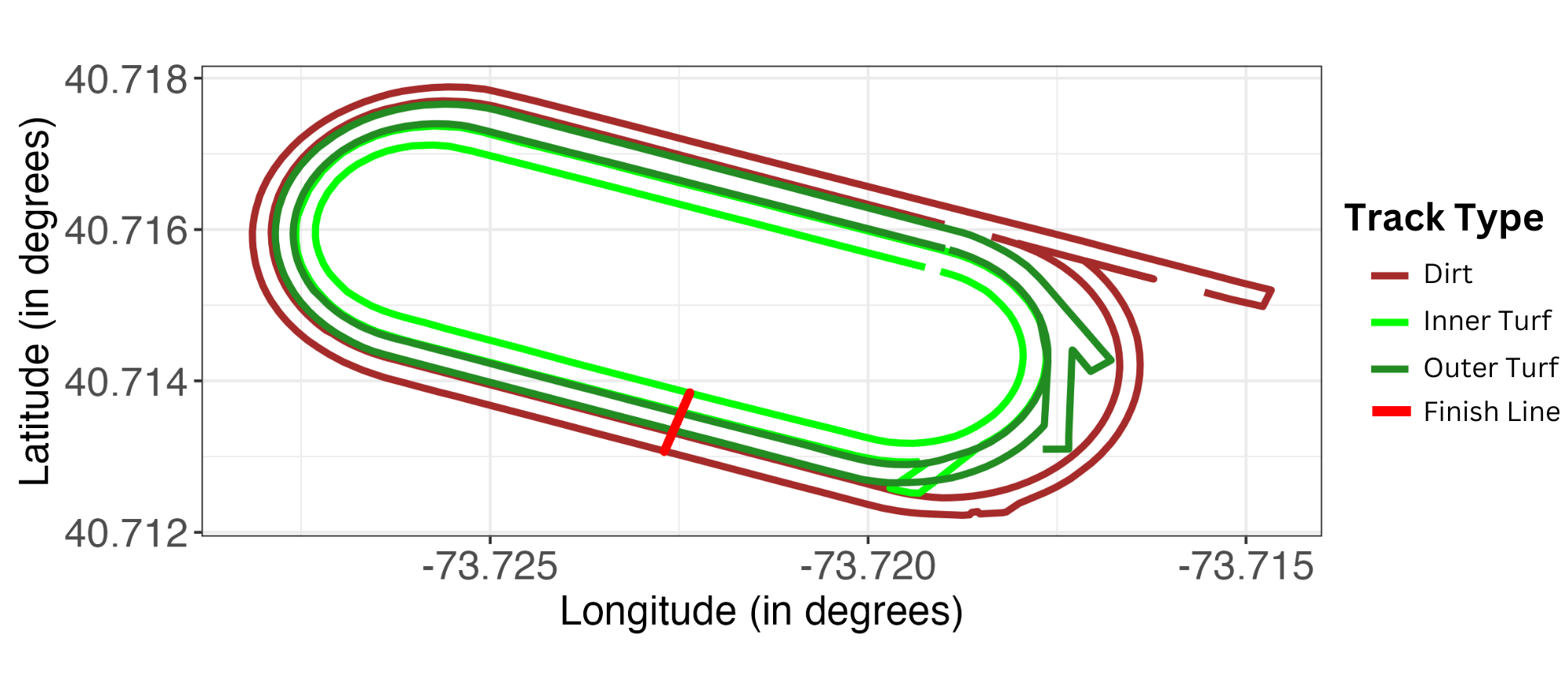}
         \caption{Raw track outlines and finish lines manually collected from Google Earth for Belmont Park.}
         \label{fig:raw_track_outlines}
     \end{subfigure}
     \hfill
     \begin{subfigure}[b]{0.45\textwidth}
         \centering
         \includegraphics[width=\textwidth]{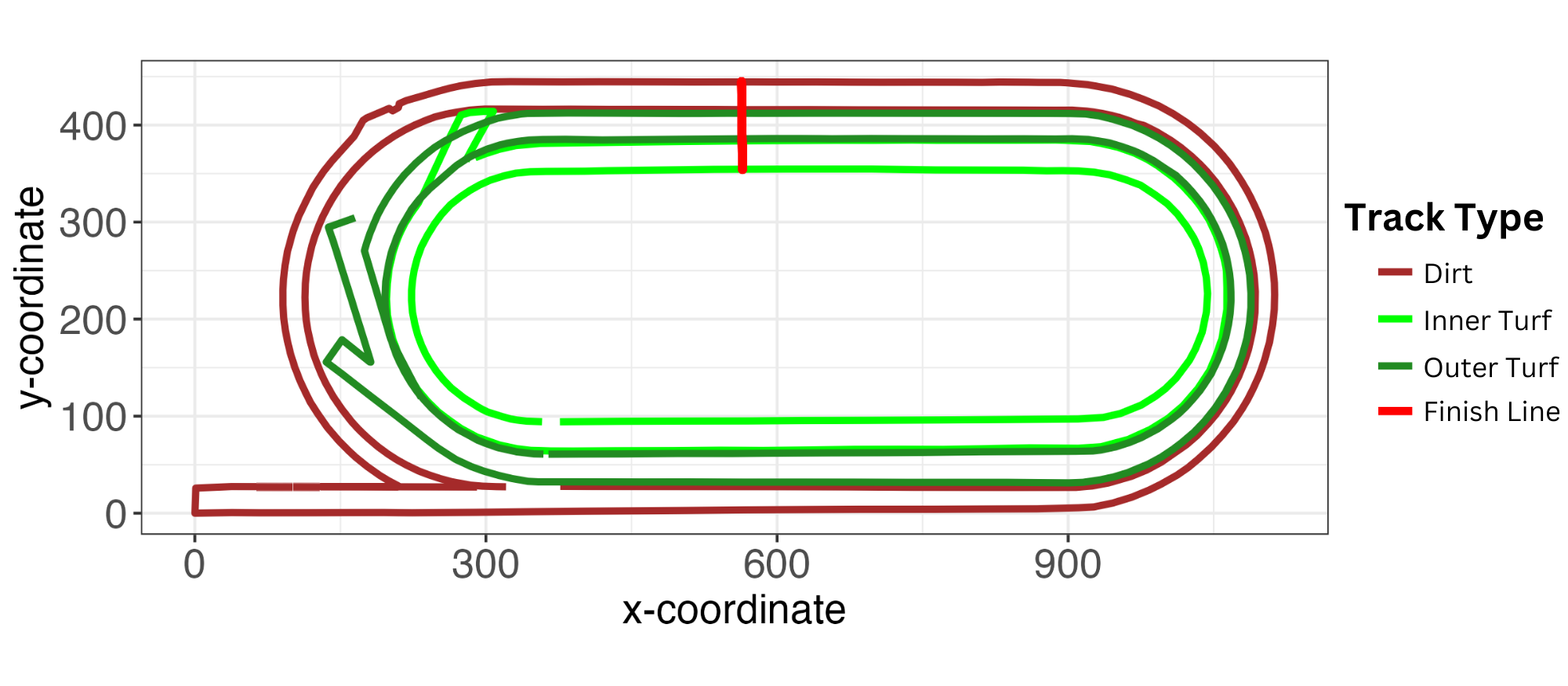}
         \caption{Transformed track outlines and finish lines from longitude/latitude to Cartesian coordinates for Belmont Park. With rotation so that the back stretch of the track is near $y = 0$ and the stretches are horizontal.}
         \label{fig:transformed_track_outlines}
     \end{subfigure}
        \caption{Track outlines for Belmont Park. This figure shows the transformation of coordinates from those collected directly from Google Earth in terms of latitude and longitude to standard Cartesian coordinates which are easier to work with.}
        \label{fig:track_outlines}
\end{figure}

The data provided by NYRA had longitude and latitude locations of the horses but did not include spatial information about the inside and outside edges of the track or the finish lines. To address this, we manually gathered track outline and finish line data for Aqueduct, Belmont Park and Saratoga using Google Earth \cite{goog}. Upon obtaining these data, we converted longitude and latitude coordinates for the track outlines, finish lines and horse location data to Cartesian coordinates using the haversine formula \citep{vanbrummelen2012} and rotating the track such that the stretches are horizontal. Figure~\ref{fig:track_outlines} provides an illustration of the raw track outlines in Figure~\ref{fig:raw_track_outlines} and the transformation to cartesian coordinates in Figure~\ref{fig:transformed_track_outlines}.\\

Upon obtaining Cartesian coordinates for the tracks, we split it into chutes, stretches and turns. Stretches are straight portions of the track, turns are curved portions, and chutes are extensions of the track used to set up the starting lanes for each horse. Chutes are manually partitioned for each track. To separate turns and stretches, we create a circle with diameter equal to the difference between the maximum and minimum y-coordinate in the inner track outline. This circle is centred at an x-coordinate equal to the minimum x-coordinate plus the radius and a y-coordinate equal to the midpoint of the maximum and minimum y-coordinate. Intuitively, the left side of the circle should trace along the left stretch of the track. We deem any portion of the track to the left of the centre of the circle to be part of the left turn. This process is repeated on the right side to identify the right turn.\\

\begin{figure}[!htb]
\centering
\begin{subfigure}[b]{.45\textwidth}
\centering
\includegraphics[width = \textwidth]{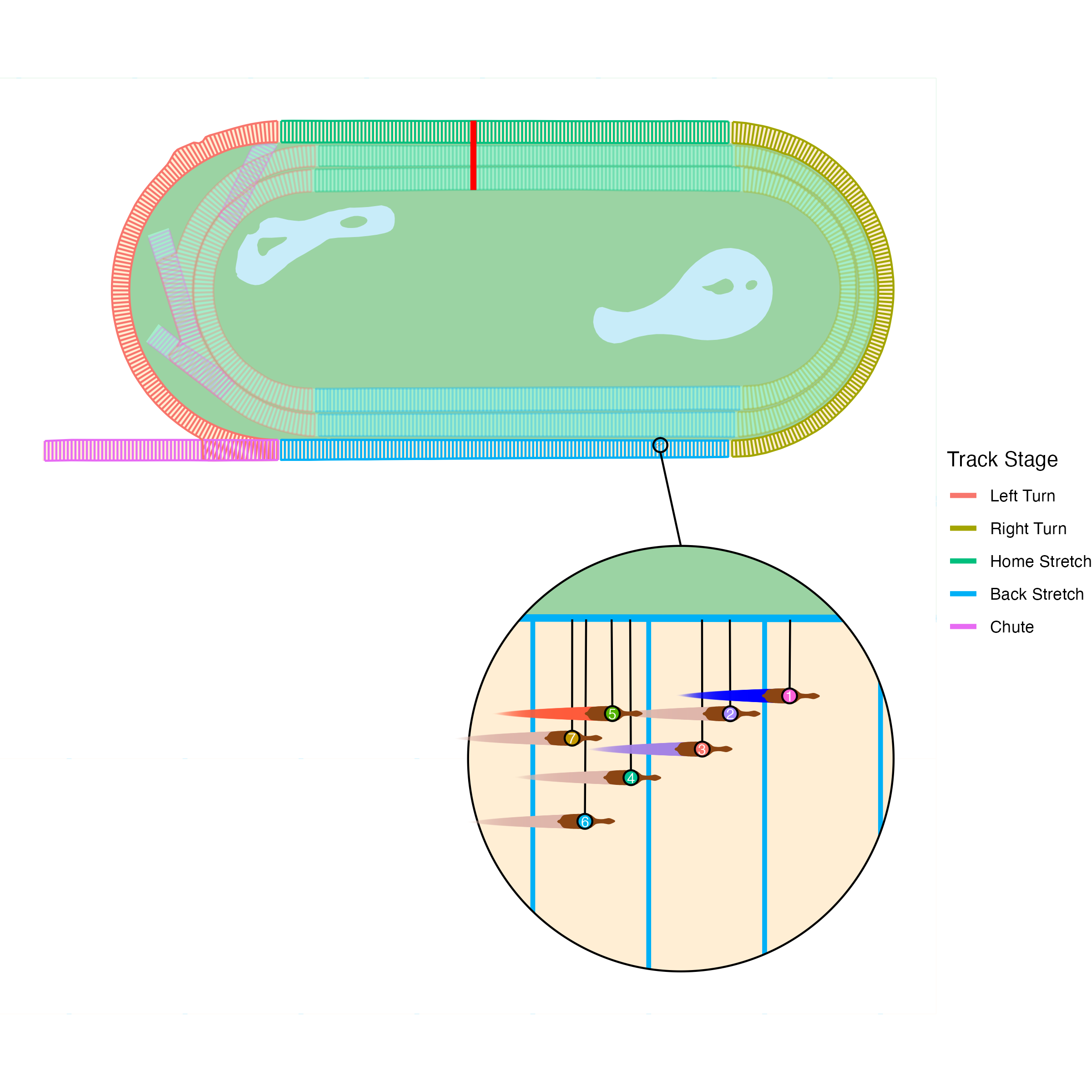}
\caption{\label{fig:final_track_outline} A snapshot of the 3rd race at Belmont Park on 2019-05-16 using our cleaned data. Stages of the track are partitioned with perpendicular lines along the track at each 10m mark. Brown shapes with coloured dots represent the horses, the coloured trail behind each horse represents its speed - with blue to red representing slow to fast - and black lines illustrate the point at which the horse is located along the inside of the track.}
\end{subfigure}\hfill
\begin{subfigure}[b]{.45\textwidth}
\centering
\includegraphics[width=\textwidth]{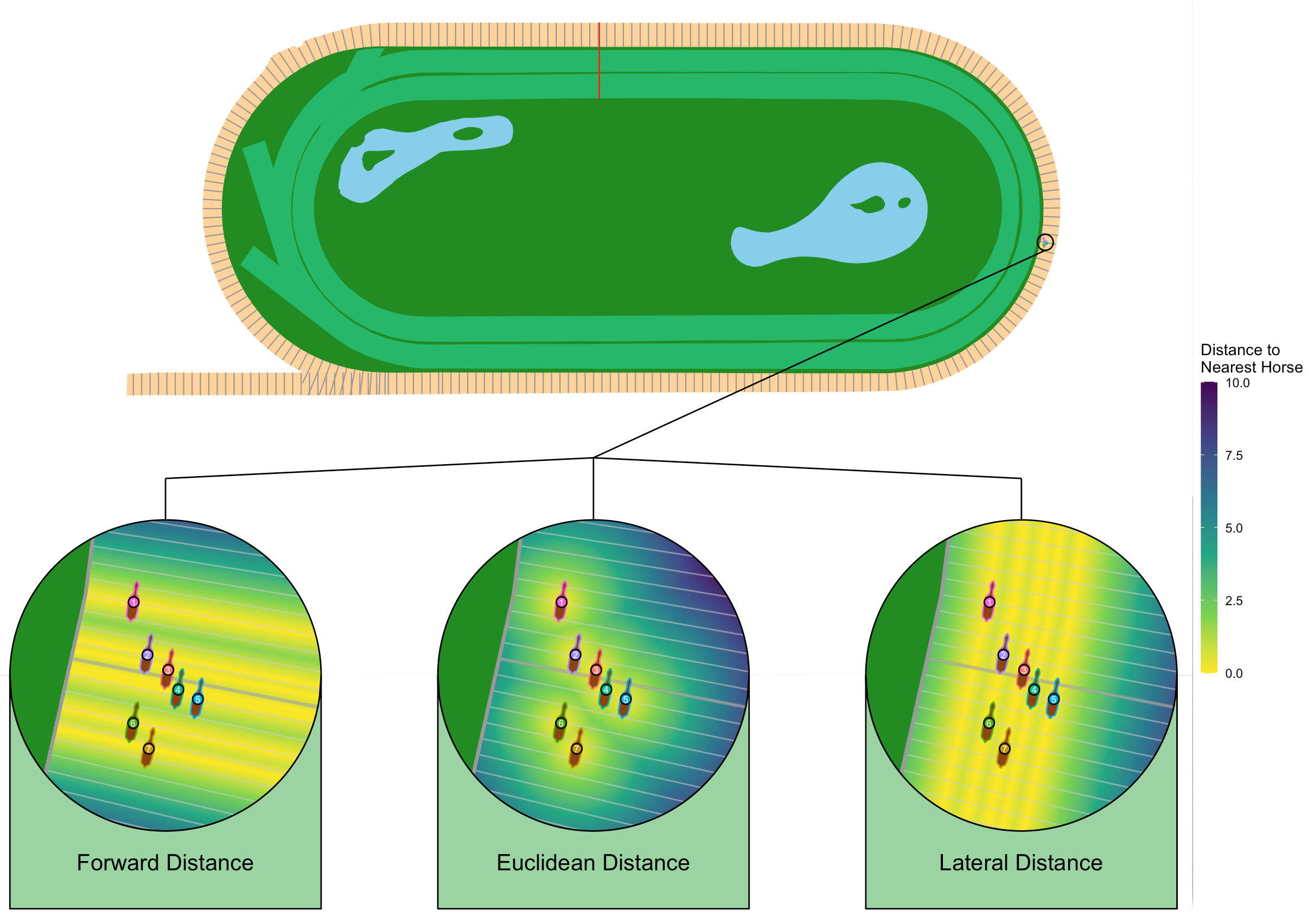}
\caption{\label{fig:distance} A snapshot of Race 3 at Belmont Park on 2019-05-16 and illustration of how several important dynamic spatial covariates are calculated. Namely we show the forward, euclidean, and lateral distance to the nearest competitor at this moment in time in the three display circles from from left to right respectively. These covariates capture a low dimensional representation of whether a competitor has space to manoeuvre in various directions and as such play a large role in our generative models.}
\end{subfigure}
\caption{Illustration of a race snapshot. At each snapshot of the race many dynamic covariates are calculated. In Figure \ref{fig:final_track_outline} we show how we partition the race track into track segments. In Figure \ref{fig:distance} we show the calculation of several important covariates relating the relative locations of the horses in the race.}
\end{figure}

Finally, we linearly interpolate at a rate of 10cm along the inside of the track. We then find the point along the inside of the track at which the distance to the horse's location is minimized for each horse. This provides us with a sense of the horse's forward location along the track, rounded to the nearest 10cm. Additionally, we take the distance between the horse and the inside of the track to be the horse's lateral positioning with respect to the inner track outline. We then use the change in forward and lateral location of the horse over each frame to describe the horse's movement on a frame-by-frame basis. Figure~\ref{fig:final_track_outline} illustrates the lateral location from the inside of the track as the length of the black lines connecting each horse to the track outline and the forward location as the point at which the black lines meet the inside of the track.\\

Note that we also use the forward, lateral, and total (Euclidean) distances between horses at each frame to create a suite of metrics that quantify a horse's positioning relative to the competition. Figure~\ref{fig:distance} illustrates the distance to the nearest horse for each of these three measurements. Using these distances travelled between frames, we are able to determine horse positions during the race. Further, we can determine the forward, lateral, and Euclidean distance between any two horses; from this we can determine if a horse is in a draft position as well as its future possible trajectories.\\

When necessary, we smooth the trajectory of a horse using an imputation based on their opponents' acceleration patterns. We sometimes observe cases in the tracking data where a horse freezes in a certain location for multiple frames then reappear improbably far down the track. This was generally an issue near the beginning of the race. Since a horse's speed is non-linear - particularly near the beginning of the race - linear interpolation would not be an appropriate solution for this issue. Instead, we leveraged information from horses that were not absent from the tracking data in those frames. If we are missing tracking data for a horse from frame $a$ to $b$, we use the average of the proportion of distance travelled between frame $a$ and $b$ by all horses with reliable tracking data. This provides us with a more realistic approximation of the horse's acceleration pattern when missing from the tracking data. We apply this imputation process to the first 40 frames (approximately 10 seconds) of the race for 4.2\% of horses across all one-mile races to stabilize the tracking data when necessary. Horses that require imputation may lack the same level of detail in the frame-by-frame positioning updates compared to their non-imputed counterparts. However, this imputation process still leverages the horse's known positioning at the starting and end moment of imputation during the acceleration phase of the race and thus the average speed over the missing interval is still correct. One can think of this imputation procedure as shrinking the missing frames towards the speed of the average horse in each interval with the constraint that the average speed over all missing frames is equal to the actual average speed travelled by the horse. Since the proportion of horses for which this issue occurred is small and the imputation procedure effectively leverages all of the information present it is unlikely this procedure has a large influence on the overall results, although it may be true that we underestimate the amount of uncertainty during some such segments. In extensions to this work, one could build such an imputation procedure directly into the joint Bayesian generative model to properly propagate the uncertainty but we leave this for future work.

\subsection{Feature Engineering}

We construct multiple features used for the novel methodology. In Section \ref{sc:method} we discussed the importance of using spatial features to represent the relationships between competitors. In this work we considered a simple representation of spatial information largely based on forward, lateral, and Euclidean distances (and position) to the nearest horse frame-by-frame. In addition, we conditioned on the number of opponents each horse is surrounded by on either side and in front during a race. In principle, one could imagine a richer set of spatial relationships but we found even simple relations captured the large majority of the variation in lateral movement in particular.\\

With this, we can construct a set of dynamic features for each horse that describes its relative position and movable space in its immediate area. This allows us to engineer a drafting model, which we describe in the next subsection. We further adjust for the effect of course type and track condition. We also generate horse and jockey effects features, which are discussed in the next section. Table \ref{tab:features} in the appendix provides a summary of the features and predictors used in our forward and lateral movement models.

\subsection{Drafting}\label{sc:drafting}
Drafting is an important factor in many multi-competitor races, including horse racing \citep{spence2012speed}, however it may be difficult to model with the appropriate level of detail without a model which operates at the granular within-race level. This may explain why, to the best of our knowledge, the literature on drafting in multi-competitor races is largely restricted to studies of aerodynamics and physics and not directly linked to performance. We believe our generative model offers a unique opportunity to study the effects of such a dynamic strategy in detail and more importantly serve as a proof of concept for future development in this area.\\

A horse, or more generally a competitor, is required to remain behind another in order to draft at all, potentially sacrificing position and/or speed in that moment. The benefit comes in the form of saved energy and potentially increased speed in the later stages of the race. What matters when deciding to draft is whether the set of race paths from that moment forward are improved or not. One also has to be careful to separate horses and jockeys particularly adapted to certain strategies from the strategies themselves.\\

To create our drafting feature, we develop a three-dimensional computer-aided design (CAD) of a horse and jockey. With this design, we use the open source software Blender \citep{blender} and OpenFOAM \citep{jasak2009openfoam} to create a 3D model of a horse and jockey, analyze the computational fluid dynamics of the model, and construct simulations. In this work, we primarily aim to demonstrate the feasibility of developing dynamic drafting covariates in a generative modelling approach at the frame-level. As such we make several simplifying assumptions regarding aerodynamics in order to generate drag coefficients which balance realism and computational efficiency. We leave improvement to fluid-dynamic modelling and more specific development on drafting models for multi-competitor races to future work.\\

One of the fundamental simplifying assumptions we made was regarding the boundary conditions for the fluid-dynamic simulations. We assume a simplified set-up wherein the horse and jockey are enclosed within a virtual closed box with air flow coming from the front. In practice there may be wind coming from other directions or other conditions which impact the flow of air and the direction of these imparted forces.  On average we expect this assumption to be reasonable and it allows us to primarily focus our attention on the dynamics regarding the horse pushing against the air in front of it and the flow of air around the horse creating pockets of lower resistance.\\

In early tests we additionally allowed for skin friction drag, as opposed to only the form drag. Skin friction drag is caused by the interaction of the friction on the surface of an object and the fluid air. In some settings, skin friction can be an important force and, in fact, NASA estimated it to be the dominant drag force for subsonic rocket applications totaling 45\% of the total drag \citep{fischer1974general}. The two most important considerations for skin friction drag are the speed of travel, since this drag is a function of the squared velocity, and the surface area interacting with the fluid. Relative to applications like flight and rocket travel, horses travel at very slow speeds and their surface area interacting with air head on is also very small which means we expect the skin friction drag to be relatively small as well. Early simulations confirmed that adding skin friction drag had nearly no effect on the simulations for horse racing and it was thus subsequently ignored, but this should be kept in mind as a potentially important force in some applications, particularly those involving high speeds and larger surface areas.\\

Given the above assumptions about the nature of the fluid dynamic forces, we additionally made two types of computational approximations. The first is regarding the mesh or cell size. In computational fluid dynamic simulations one is estimating the solution to a set of Navier-Stokes equations. To aid in computation the object of interest is split into small pieces by a mesh and the equations are solved individually on each piece and added back together linearly to determine the result of the simulation on the total object. In our case, the simulation on each cell was solved using the pressure-implicit with splitting operators (PISO) algorithm provided by OpenFOAM. As the mesh becomes finer and we break the object of interest into smaller pieces the fidelity of the simulation increases at the expense of computation. In sophisticated mesh designs one can divide the object of interest unevenly across regions with different features. In our case, the body of the horse acts as a solid rectangle against the incoming air, whereas the face is curved and interacts with the air in less trivial ways. These special areas with the highest local curvature were given a finer mesh as we expect the simulation results vary more greatly across these areas. To determine the final mesh, we used an iteration method recommended by \cite{simscaleWhatWhat}. An initial mesh was chosen by visual inspection taking into account the regions of high and low local curvature as discussed above. Then a series of simulations were conducted with each simulation using a finer mesh than the previous. The iterations continued until the results converged with respect to a pre-specified tolerance, which we chose in this case to be a 5\% relative change in simulation results. In some applications, a smaller tolerance will be desirable to choose. Additionally, we checked the estimated $y+$ value from the final simulations as provided by OpenFOAM to ensure it met the diagnostic condition appropriate for this class of simulations ($30 < y+ < 300$)\citep{simscaleWhatyplus}. The $y+$ measure is important for determining the simulation performance near the boundaries or walls in computational fluid dynamics simulations.\\

The second kind of numerical approximation which was necessary was with respect to the distance between horses. The drag reduction from drafting is a function of distance to the horse in front. As that distance increases the relative decrease in drag experienced by the trailing horse goes to zero. Similarly being directly behind a horse will decrease the drag experienced more than being slightly to the left or the right of the horse in front. The goal of our simulations is to be able to return a coefficient of drag as a function of any relative positioning of the trailing and leading horse, however we are unable to run full simulations at all possible (continuous) distance values. To approximate the drag function we specified a two-dimensional grid of distances where we ran simulations and then linearly interpolated to generate the drag coefficients used as a basis for covariates in the various distance models. We ran simulations on a $3\times 3$ grid of locations at which the drafting horse is located where the drafting horse is either 2, 3.5, or 5 meters behind and either directly behind or 0.5 meters to the right or left. Finally, we also calculated the drag in the scenario of no drafting or otherwise called the clean air condition.\\

\begin{figure}
     \centering
     \begin{subfigure}[b]{0.47\textwidth}
         \centering
         \includegraphics[width=\textwidth]{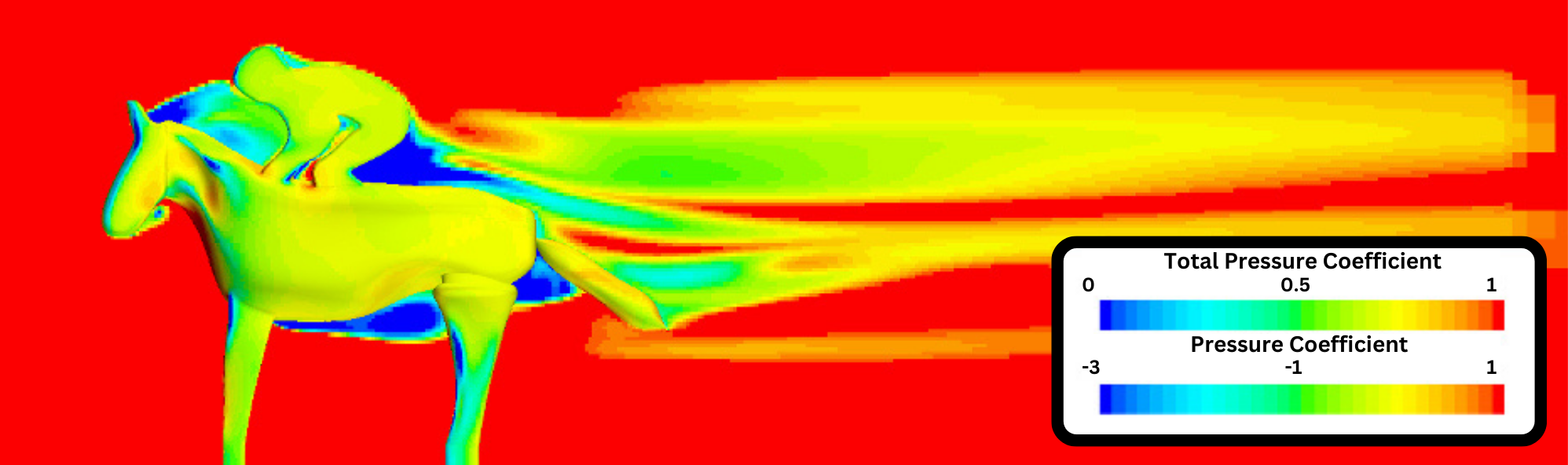}
         \caption{Measurements of air pressure on a horse and jockey in clean air.}
         \label{fig:draft1}
     \end{subfigure}
     \hfill
     \begin{subfigure}[b]{0.47\textwidth}
         \centering
         \includegraphics[width=\textwidth]{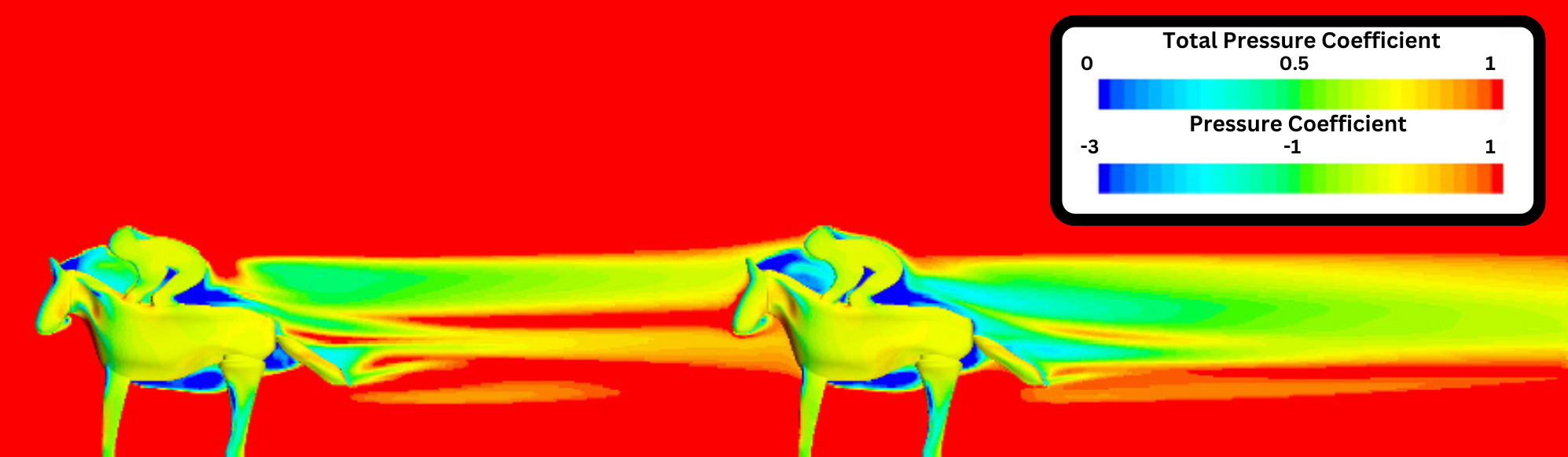}
         \caption{Measurements of air pressure for two horse and jockey pairs, with one drafting behind the other.}
         \label{fig:draft2}
     \end{subfigure}
     \hfill
     \begin{subfigure}[b]{0.55\textwidth}
         \centering
         \includegraphics[width=\textwidth]{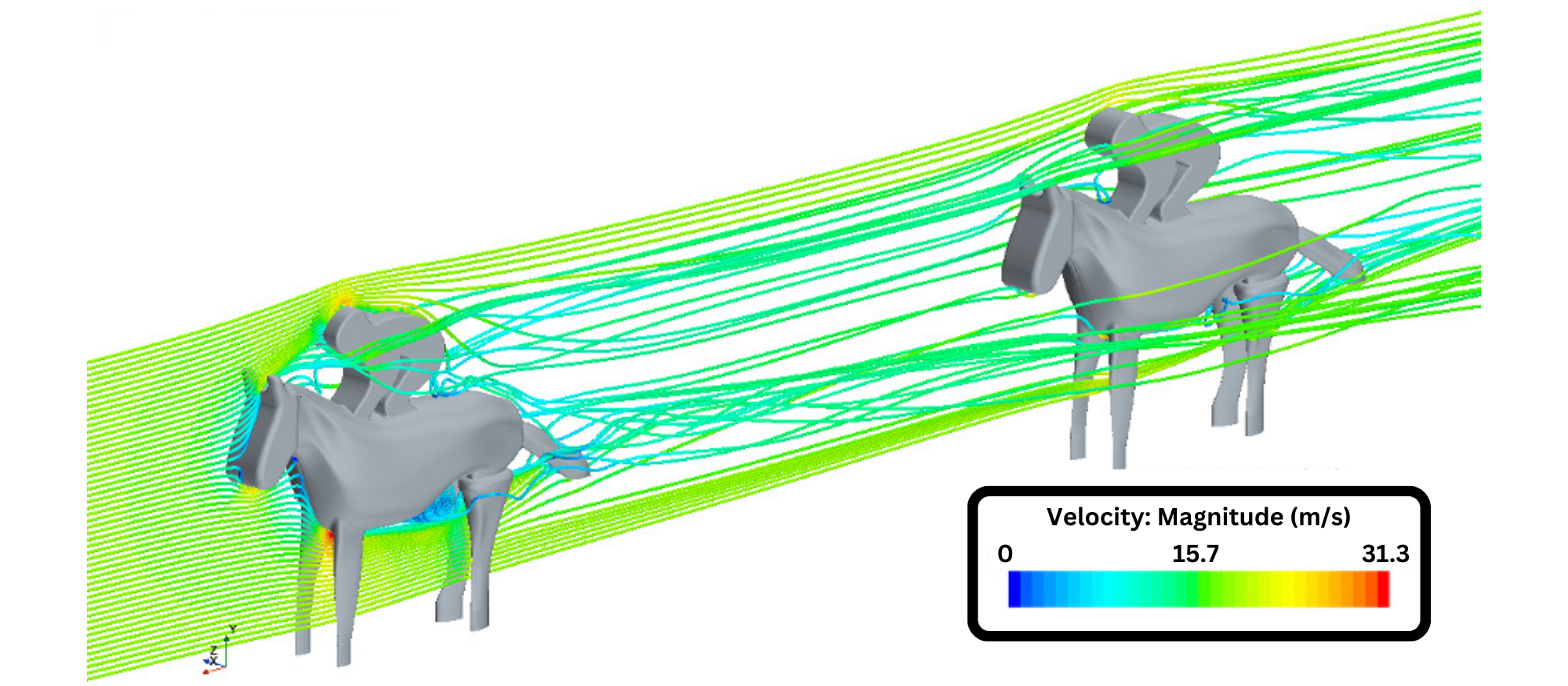}
         \caption{A visualization of simulations of the fluid dynamics of two horse and jockey pairs, with one drafting behind the other. Obtained from applying the Pressure-Implicit Splitting Operators (PISO) algorithm with OpenFOAM.}
         \label{fig:draft3}
     \end{subfigure}
        \caption{An illustration of our drafting model and simulations. In Figure \ref{fig:draft1} we show a representation of the pressure coefficients generated from a fluid dynamics simulation in clean air. In the subsequent Figure \ref{fig:draft2} we see a visual representation of the drag coefficients from a fluid dynamics simulation with two horses, one trailing another or drafting. In Figure \ref{fig:draft3} we see an illustration of the drag effect. That is, we can see that the trailing horse is subject to less air resistance according to the fluid dynamic simulations.}
        \label{fig:draft}
\end{figure}

Using the estimated grid of drag coefficients from the simulations and the linear interpolations thereof we created two types of covariates which we then used in our models. The first covariate was a simple indicator of whether a horse was currently drafting, or currently located such that they were benefiting from a reduction in air drag, or not. Second, we estimated the total proportion of energy saved from drafting up to each moment in the race. This is done by calculating the energy, $E$, used by a horse as $E = F_d \, s$, where $F_d = \frac{1}{2}\,\rho\,{v}^2\, c_{d}\,A$ is the force experienced by the horse while trying to move forward, $\rho$ is the mass density of air, $v$ is the velocity of the horse, $c_d$ is the calculated drag coefficient, $A$ is the frontal area of the horse (assumed to be one square metre), and $s$ is the distance covered by the horse in a frame. These two approaches allow us to capture both short- and long-term dynamics relating to drafting. In principle, much more complicated drafting simulations and covariates are possible. And perhaps more importantly one could specify relevant interactions between these covariate values and the current stage of the race.  In some multi-competitor races, it may be especially important to understand drafting dynamics in packs or groups for example. Cycling is a good example where much of the race occurs in a pack where there is much more drag reduction. In this application, however, we assume that the bulk of the drafting effect is the result of the nearest horse in front in order to limit the complexity and scope of the computational simulations required. While these assumptions may be simplifying, we believe that they represent a meaningful step toward capturing these complicated dynamics and serve as proof of concept. See Figure~\ref{fig:draft} for visualizations of the fluid dynamics and drafting simulation procedure.\\

\subsection{Model and Simulations}\label{sc:mod and sim}

To build our model, we only include the one-mile races from the 2019 season provide by NYRA and NYTHA. Our methodology can easily be extended to races of differing lengths in a hierarchical scheme, and we make this choice for both demonstrative purposes and computational efficiency. Following the general methodology outlined in Section \ref{sc:method} we develop two models, one for estimating a horses forward movement at each frame and another estimating lateral movement. Based on the data we made several simplifications to the general joint density in equation \eqref{eq:gen mod}. First we assumed independence between the forward and lateral movements. This is of course not true. In fact, there must on some level be dependence since horses only have a finite amount of energy to expend, and maximal exertion perpendicular, for example, would result in restrictions to how much lateral movement would be possible.\\ 

More formally, the most natural way one would model the dependence between the forward and lateral movement would be to make one model conditional on the other. Explicitly, one could modify the joint distribution in equation \eqref{eq:gen mod} as follows:

\begin{align}
p(\yf_i|\tf(j),\Xf_{i}, \psi_f) \times p(\yl_i|\yf_i,\tl(j),\Xl_{i}, \psi_l), \quad i = 1,2,\dots,\I,\label{joint decomp}
\end{align}
where we condition directly on the forward distance in the lateral movement moment. This requires us to simulate the forward distance first and subsequently the lateral distance which can be more cumbersome. In some cases a multivariate normal on the outcome or some transformation thereof may also be appropriate.\\

In practice with respect to our canonical example, however, since the frames are approximately 0.25s long the large majority of this effect is captured in the latent effects, the spatial information from the previous time-step and covariate information about where the horse is on the track, such as whether they are on a bend or not, and thus this simplification seemed not to impact inferences very much. More formally, when the time between frames is relatively small, i.e $\Delta t < \epsilon$, then we would expect the information at time $t$ represented by the filtration $\mathcal{F}_t$ to be well approximated by the information at the previous time step, $\mathcal{F}_{t - \Delta t}$. In the small $\Delta t$ setting, if our models are capturing the total information available at each time step, then we might expect the loss of not modelling the dependence to be more minimal since our models are explicitly conditioned on all the information available up to $\mathcal{F}_{t - \Delta t}$. In the example of the lateral model, we use the information of lateral movement in the previous frame which is highly related to the forward movement in the previous frame which is highly correlated with the forward movement in the current frame. The independence simplification, of course, may not be appropriate for all multi-competitor races, particularly those where frames are spaced out further in time where approximating the current information with that available in the previous frame may not be as credible.\\

Second, for similar reasons, we assumed that there are no latent time-varying horse effects in the lateral movement model, but instead time constant jockey effects. Preliminary testing found that over 99\% of the variation in lateral movement could be explained by simple spatial covariates, a constant jockey effect, track phase indicators which include this like turn and home stretch, and the motion from the previous time frame. Since this simple model accounted for much of the variation, the model was simplified for computational reasons. Of course, when appropriate, these effects could be made more complicated. We also assumed that horse speeds only depend on each other through the spatial covariates such as distances to nearest horses at each frame.\\

For the forward movement model we modelled the horse speed profiles with b-splines \citep{de1978practical, dierckx1995curve}. This spline technique encodes the knowledge that a horse's average speed at any point in a race is likely to be smooth without assuming too much about what that function looks like exactly and using the data to best decide. The splines are fitted using all tracks, and the knot placements for the splines were decided both using a leave-one-out cross-validation approximation and a visual assessment. The knot placements correspond roughly to strategy transitions, for instance the end of the initial acceleration at the start of the race as well as the final quarter-mile. The b-splines were generated with the splines2 R package \citep{wang2021shape}.\\

Overall, the forward model for each competitor k looks like:
\begin{align}
\yf_i(k) \sim N(\tf_k(j) + \delta^f_{jockey} + \delta^f_{track} + \bx^{\text{for}}\psi_x, \sigma_f),
\end{align}
where $\tf_k(j)$ is the $k$th competitors spline value at location $j$, the $\delta$ parameters represent track and jockey effects and $\bx^{\text{for}}$ represents all other covariates which are listed in Table \ref{tab:features}.\\

The finite vector of spline parameters and the track and jockey effects were all regularized using random effects structures of the form:
\begin{align}
\delta \sim n(\mu_{\delta}, \sigma_{\delta}),
\end{align}
where $\mu_{\delta}$ was treated as an unknown mean for the spline effects, but fixed at zero for the jockey and track effects and $\sigma_{\delta}$ was a fixed hyper-parameter for all three parameter types. Thus the spline parameters were shrunk towards the average speed for that portion of the race and jockey and track effects were shrunk towards 0. Covariate and outcome variance coefficients were given weakly informative priors.\\ 

When fitting the forward model on all the data, particularly in the model exploration phase, we used the optimization functions in the RStan package \citep{carpenter2017stan}. Optimization generates an approximately Bayesian model via Maximum A Posteriori (MAP) estimates. In principle, one could allow the random effect variance parameters to be unknown, for example, but this may be more suitable for variational or MCMC methods with the trade-off being longer compute times. We found that the posterior means and posterior predictive means tended to be well-behaved using Optimization, but that occasionally these fits produced poorly behaved tails and subsequently unrealistic simulations. Additionally, when using truncated distributions the optimization approach sometimes failed to converge. When generating later simulations on a handful of horses in Section \ref{fig:cf sim} we fit full MCMC models on a subset of the data and additionally truncated the forward model below at 0. When the goal was generating expectations, truncating or not did not have large impacts in most cases, but if the object of interest was realistic and interpretable simulations we found truncating to be important.\\

The second model we construct is a lateral movement (LM) model. This models the lateral speed of each horse. Throughout this work we use the terms side movement and lateral movement interchangeably. For this model, we again use a simple Gaussian model:
\begin{align}
\yl \sim N(PLM\beta_{plm} +\delta^l_{jockey} + \delta^l_{track} + \bx^{\text{lat}}\psi_{lat}, \sigma_{l}) ),
\end{align}
where again the track and jockey effects, $\delta^l_{track}$ and $\delta^l_{jockey}$, had random effect structures and the lateral covariates found in Table \ref{tab:features} had weakly informative priors scaled to movement speeds possible for a horse. The most important covariate in this model is $PLM$ which is the horse's previous lateral movement from the past frame. This encodes the fact that horses moving to the inside or the outside tend to continue doing so since the frames are so close in time.\\

\begin{figure}
\centering
\begin{subfigure}[b]
{0.45\textwidth}
\centering
\includegraphics[width=\textwidth]{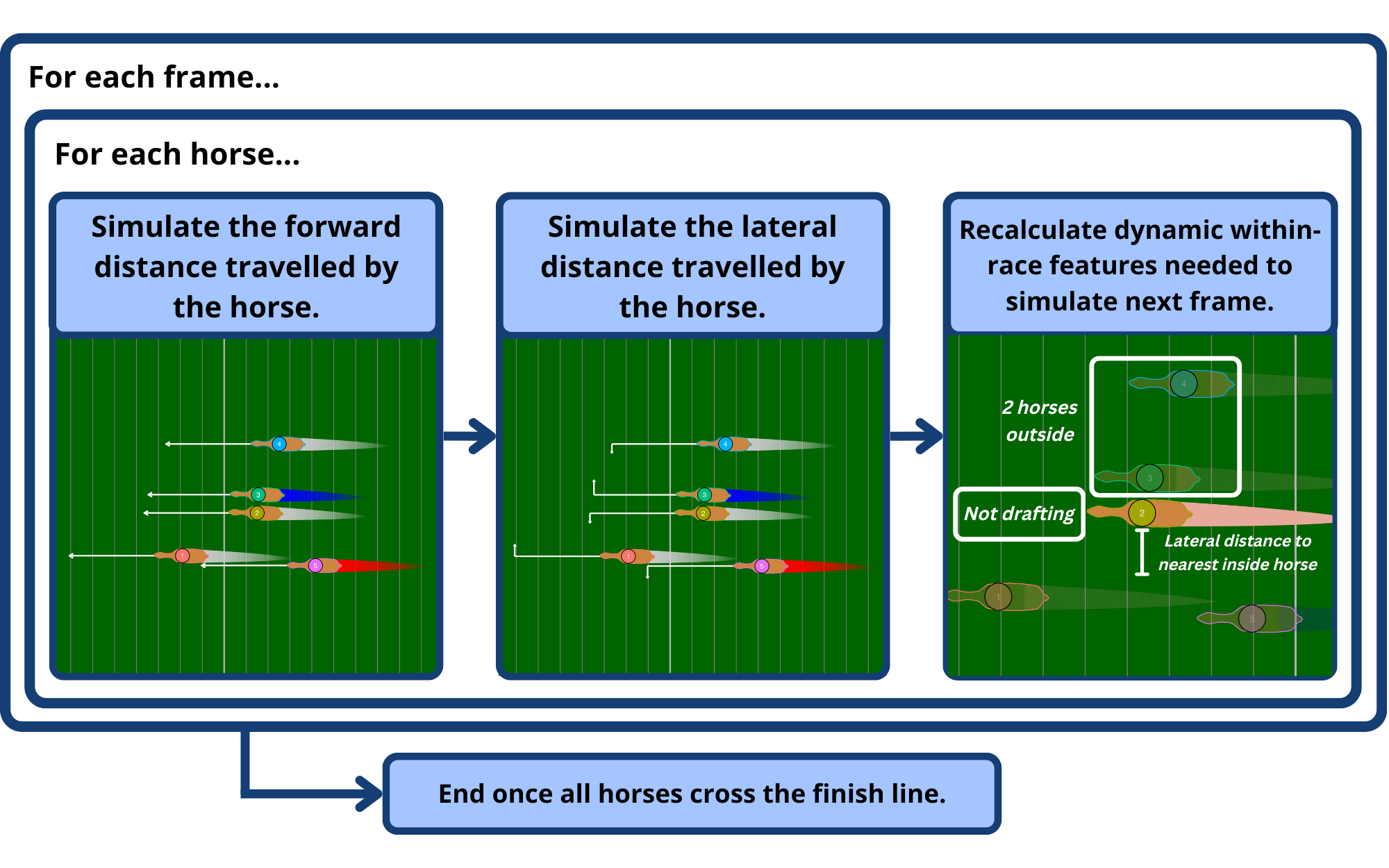}
\caption{\label{fig:model_sum} Simulation procedure leveraging the forward and lateral movement models. This illustration is meant to provide a high-level pseudo-algorithm for the prediction of horse races using this model framework. For each horse, we 1) simulate the forward distance travelled in the next frame, 2) simulate the lateral distance travelled in the next frame, 3) after applying steps (1) and (2), update the spatiotemporal features. This process is repeated on a frame-by-frame basis until all horses cross the finish line. Note that lateral movements are exaggerated for illustrative purposes.}
\end{subfigure}
\hfill{1}
\begin{subfigure}[b]{.45\textwidth}
    \centering
\includegraphics[width=\textwidth]{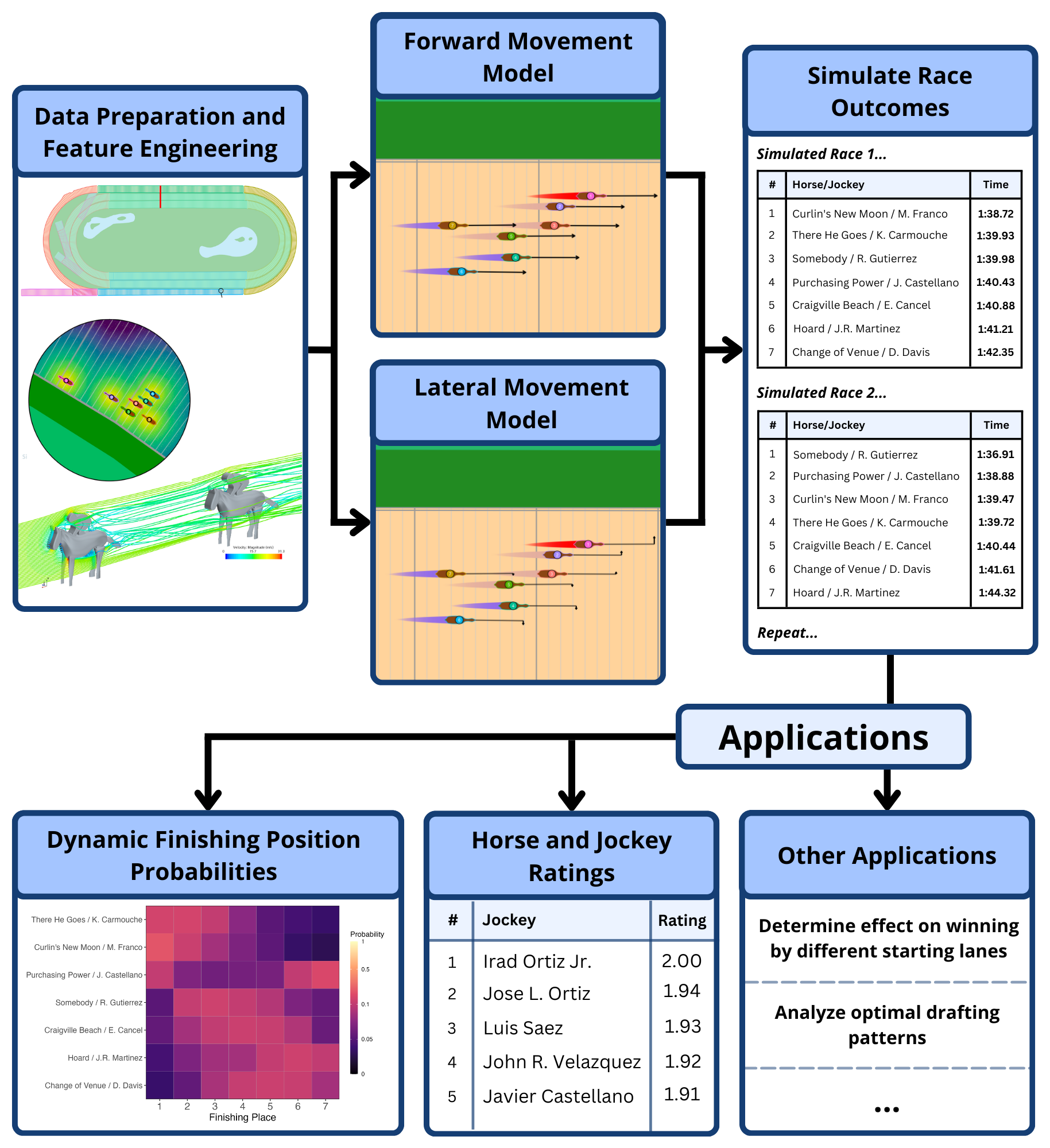}
\caption{\label{fig:model} Full methodology summary. We begin with data preparation and feature engineering such as includes extracting track outlines from Google Earth, smoothing tracking data via imputation and calculating distance features based on the relative positioning of the horses. Next, we fit approximate Bayesian models to predict the forward and lateral movement of each horse on a frame-by-frame basis. We then use those models to predict race outcomes at any moment of a race by running thousands of simulations of the forward and lateral movement of horses for each frame until all horses have completed the race. This simulation process as well as the model parameters and effects provide a wide range of applications in multicompetitor races with spatiotemporal tracking data.}
\end{subfigure}
\caption{Illustration of the simulation procedure and full modelling pipeline, respectively. These flowcharts represent the step-by-step process taken in building this project.}
\end{figure}

The forward and lateral models give us a straight-forward method for simulating entire races. We simply iterate between simulating forward motion and then lateral movement for all horses simultaneously frame-by-frame ensuring we save the current location of all horses and calculate all dynamic covariates at each step. See Figure~\ref{fig:model_sum} for a summary of the simulation algorithm.\\

While simple, simulating over a sufficient number of posterior draws may be computationally cumbersome even for a single race. One-mile races, for example, tend to last approximately 100 seconds which generates on the order of 400 frames in this data set. Simulating over 2000 posterior draws with 6 competitors requires $6\times 2\times 400 \times 2000 = 9.6 \times 10^6$ draws from normal distributions in addition to updating the positions and recalculating the dynamic covariates. This is largely infeasible at scale in standard R coding. We were able to make the problem feasible leveraging the fact that Stan is written in C++. Rstan \cite{carpenter2017stan} has a function $gqs(\cdot)$ which gives direct access to the generative quantities block. This allows us to write the posterior simulations directly in this block and execute them both separate from fitting the model but also in such a way which lends itself well to parallelization of the simulations at the race-level. Using the $gqs$ function, 2000 simulations of a race can be fit on the order of 90-120 seconds on a MacBook Pro with 64 GB of RAM and 16 cores. Early versions of this code written in standard R took 5-10 minutes to complete fewer than 50 full simulations. When saving only final outputs or summaries of the simulations the memory load is significantly lower and these computation times can be reduced significantly in many cases.\\

As discussed, these simulations allow the computation of various notions of instantaneous value. For example, in Figure \ref{fig:model} we can see dynamic placing probabilities for all horses at a given snapshot of a real race. In the following sections we will discuss some of the inferences generated from this model as well as some examples of the kinds of analysis a fully generative multi-competitor race model is capable of.

\subsection{Model Evaluation}

At a high-level our recommended philosophy for model evaluation 
of these kinds of models incorporates three elements. First, we want to understand how well our models operate on the data granularity level on which the data is trained. Since these are frame-level models, we evaluate their performance in terms of how well they predict frame level outcomes (in and out of sample) such as say the actual forward or lateral distance travelled in a frame. Second, we want to look at outcomes on a larger more meaningful unit. In our canonical example we looked at race-level outcomes. Race-times and rankings are examples of units for which we can generate predictions but for which our model is not directly trained and for which different modelling choices can be evaluated. In some multi-competitor races there may be additional sub-units below the race-level for which such evaluation is also valuable. Finally we looked at the inferred coefficients and the produced simulations. Poor modelling choices can lead to more unrealistic race paths and latent values which strongly contradict domain expert knowledge. This step can most benefit from collaboration with experts.\\

Since the models are computationally very expensive we also took advantage of using simplified models for various parts of the model evaluation process. In some cases this meant simplifying the number of covariates to only the most important ones or fitting on a smaller subset of horses or races to test out simulations. Building out these models in layers of complexity can be crucial both from a development time standpoint and in understanding how additional features or more complicated model architectures change the predictions and inferences generated.\\

In Section \ref{sc: profile} we describe how we chose the hyper-parameters relating to the forward speed profiles such as the number of knots and their placement with respect to the above outlined evaluation principles. These choices were found to be the most important modeling decisions with respect to overall performance.

\section{Results}

\subsection{Race Simulation and Dynamic Win Probability}

\begin{figure}
     \centering
     \begin{subfigure}[b]{0.47\textwidth}
         \centering
         \includegraphics[width=\textwidth]{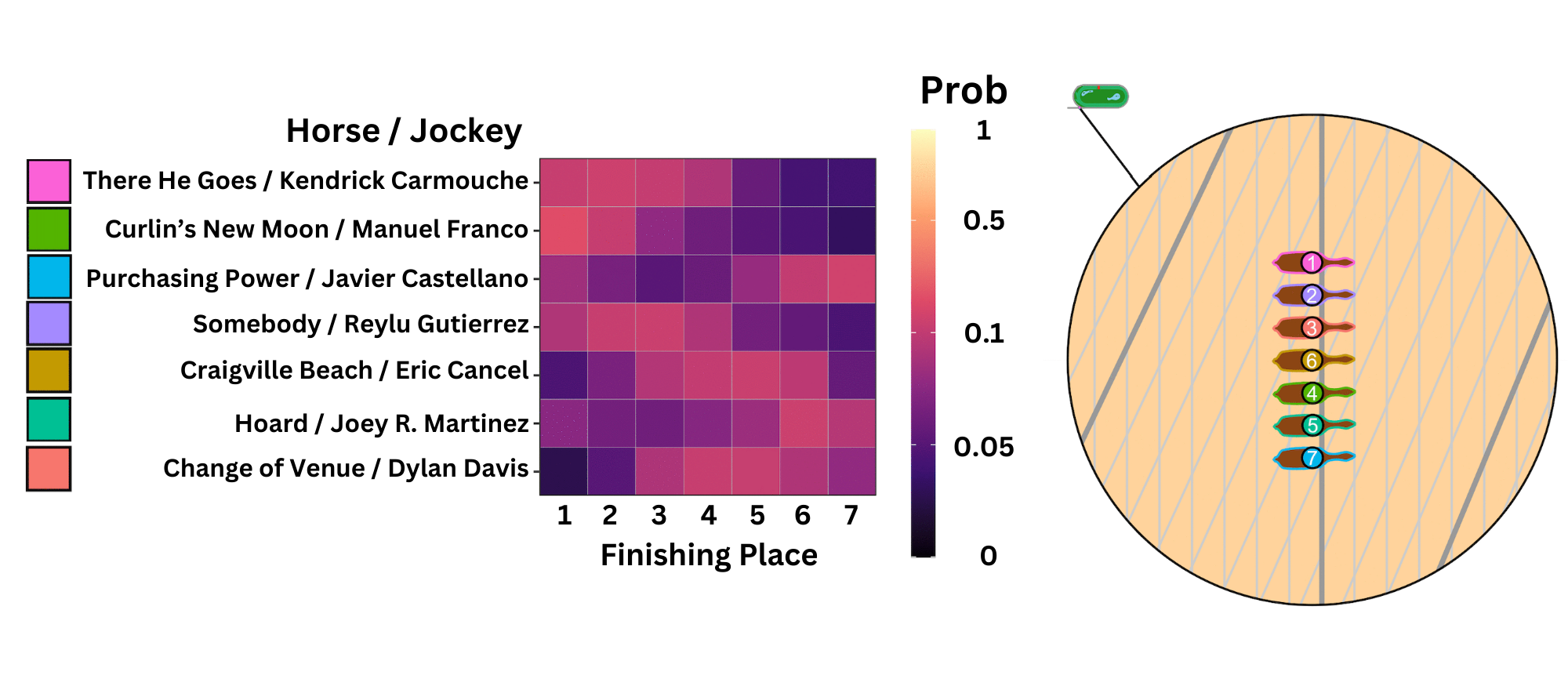}
         \caption{Placement probabilities of all horse/jockey combinations at the beginning of the race. At the beginning of the race, horse spline effects, jockey effects and starting lanes are the primary influencers on placement probabilities. Purchasing Power (blue) is at a disadvantage starting on the outside lane. Whereas, There He Goes and Somebody are in the more advantageous inside lanes.}
         \label{fig:dynamic_win_prob_start}
     \end{subfigure}
     \hfill
     \begin{subfigure}[b]{0.47\textwidth}
         \centering
         \includegraphics[width=\textwidth]{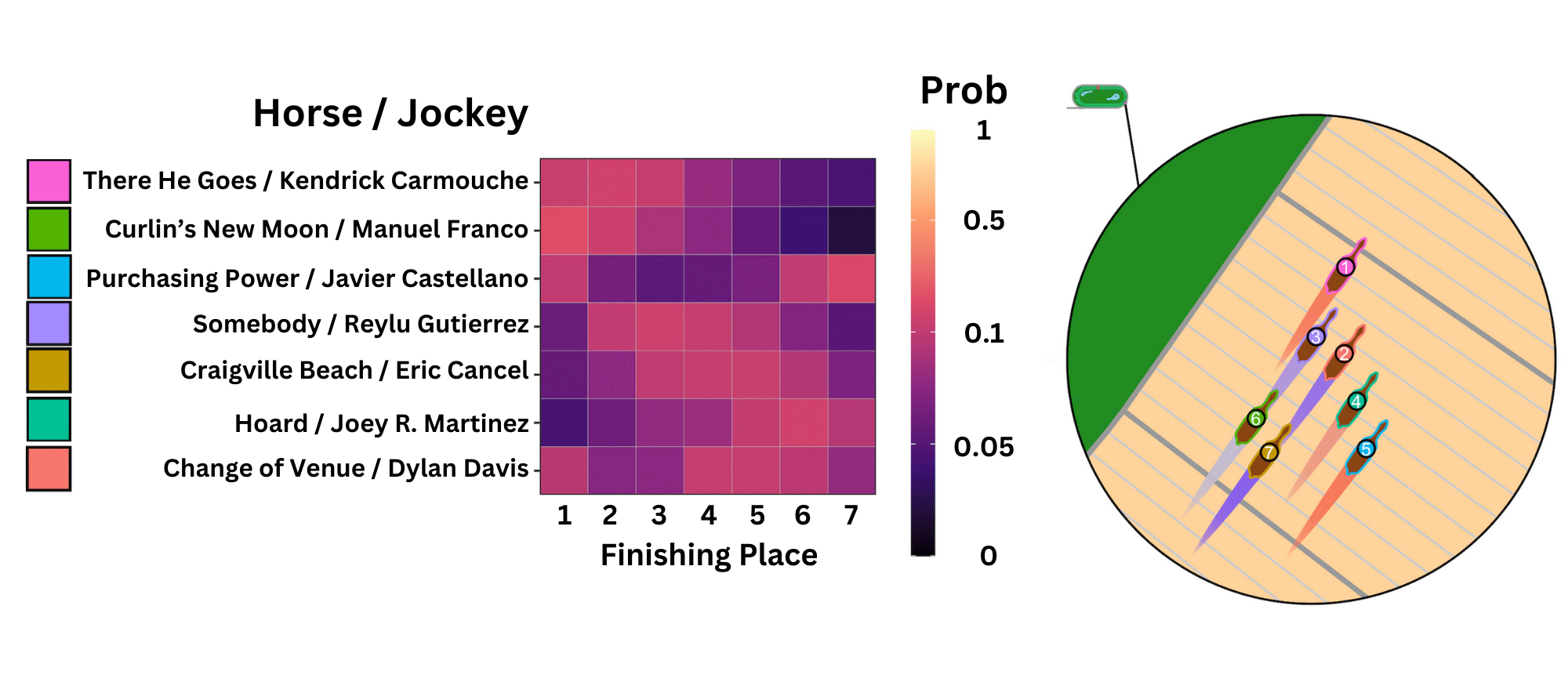}
         \caption{Placement probabilities of all horse/jockey combinations at the half way point (50.2 seconds since start). There He Goes remains one of the favourites to win at the front of the pack while Purchasing Power remains the most likely candidate to finish 6th as he remains on the outside lane.}
         \label{fig:dynamic_win_prob_middle}
     \end{subfigure}
     \hfill
     \begin{subfigure}[b]{0.47\textwidth}
         \centering
         \includegraphics[width=\textwidth]{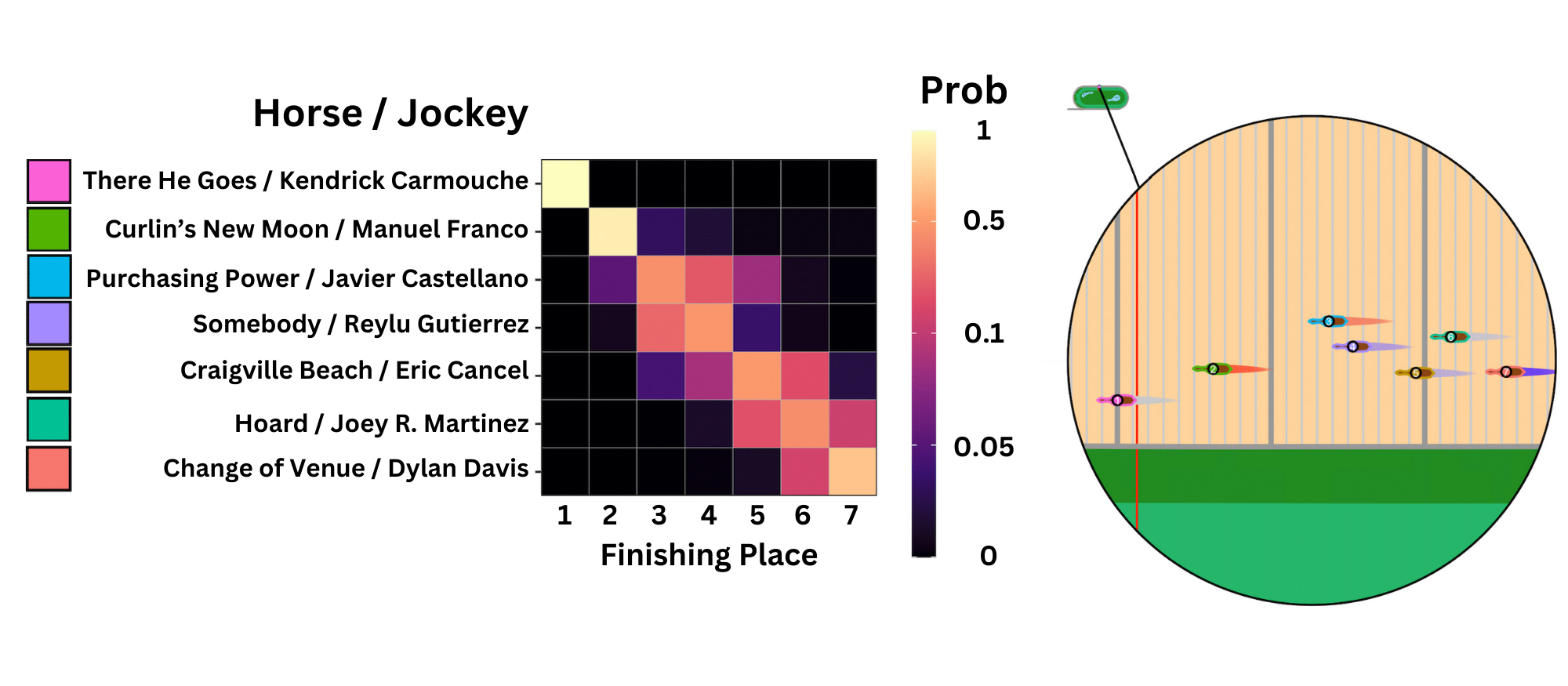}
         \caption{Placement probabilities of all horse/jockey combinations as the 1st place horse crosses the finish line. As the end of the race nears, the placement probabilities begin to converge to each horse's true finishing placement. However, uncertainty still remains as some horses are very tightly placed such as Purchasing Power and Somebody in 3rd and 4th, respectively.}
         \label{fig:dynamic_win_prob_end}
     \end{subfigure}
        \caption{Illustrative example of dynamic win probability behaviour using data from the 3rd one-mile race at Belmont Park on 2019-05-16. The square heatmap on the left represents the placement probabilities for all horses on the y-axis and all finishing placements on the x-axis. The visual on the right displays the horse positionings on the track at the corresponding frame where horse colours correspond to the rainbow colour scale to the left of the horse and jockey names.}
        \label{fig:dynamic_win_prob}
\end{figure}

At each race frame, our model performs multiple simulations that predict the remainder of the race. Using the extracted finishing position of horses in these simulations, we are able to construct dynamic race finishing placement probabilities for each horse. As more information is provided with each frame that passes, these predictions become more accurate and converge to the true race result. We present how our dynamic win probabilities behave using race three from the Belmont which occurred May 16, 2019 in Figure~\ref{fig:dynamic_win_prob}.\\

In the heat maps depicted in Figures \ref{fig:dynamic_win_prob_start} to \ref{fig:dynamic_win_prob_end}, the left-most column corresponds to the probability a horse finishes in first, and the right-most column corresponds to a horse finishing last. The circular visual to the right of the heat map is a bird's eye view of the race state at that given time. As the true race progresses, the win probabilities are being updated given the new information of the race characteristics until the race finishes and the true finish order is determined. Analyzing this visualization we see that the horse Curlin's New Moon is sitting back in the first half and in a draft position. Despite sitting around fifth and seventh place, our model still predicts that this horse will have a top finish placement during the early stages of the race. Curlin's New Moon ultimately finished second this race, demonstrating our model's capability to capture effects that would otherwise be lost to the human eye. This effect is likely due to a combination of Curlin's New Moon having relative high predicted speed in the upcoming stretches of the race and his favorable positioning with respect to drafting. We can construct full simulations and thus visualizations for any mile-long race in the originally provided data set or even for races which did not occur as long as the horses are in the data set.

\subsection{Jockey Ratings}

From the forward model, we can compute the posterior mean of the jockey's random effect to produce a jockey rankings measure. The ability to compute this demonstrates the benefit of our choice to model in a Bayesian framework as well as our methodology's flexibility. This measure quantitatively describes the positive impact that a jockey has on their horse's estimated final position (i.e. the higher the rating, the greater the positive impact on race result). Table~\ref{tab:jockey} displays the top ten jockey ratings produced by our modelling procedure. We compare our ratings to the total earnings leaderboard in 2022 for Saratoga \cite{sar}, Belmont Park \cite{bel}, and Aqueduct \cite{aqu} for our model's top ten jockeys. Unfortunately, we are unable to obtain the earnings rankings from 2019 as they are not available on the track websites. However, we find that Irad Ortiz Jr., the top rated jockey from our model, also ranks first at Saratoga and Belmont, and fourth at Aqueduct. The models top jockeys have performed reasonably well on at least one of the selected tracks in 2022, 3 years post our training set, with the exception of Joe Bravo who did not compete on any of the three tracks. However, Joe Bravo has still achieved 54 first place finishes in 2022 by the end of October 2022. This suggests our model has some positive signal in identifying top jockeys in a forward predictive sense.

\begin{center}
\begin{table}[]
\centering
\begin{tabular}{cc|ccc}
\hline
Jockey            & Model Rank & Saratoga Rank & Belmont Rank & Aqueduct Rank \\ \hline
Irad Ortiz Jr.    & 1          & 1             & 1            & 4             \\
Jose L. Ortiz     & 2          & 5             & 7            & 6             \\
Luis Saez         & 3          & 3             & 6            & NR            \\
John R. Velazquez & 4          & 8             & 14           & NR            \\
Javier Castellano & 5          & 7             & 10           & 2             \\
Manuel Franco     & 6          & 6             & 3            & 1             \\
Jose Lezcano      & 7          & 9             & 8            & NR            \\
Joe Bravo         & 8          & DNC           & DNC          & DNC           \\
Joel Rosario      & 9          & 4             & 4            & NR            \\
Junior Alvarado   & 10         & 12            & 22           & NR            \\ \hline
\end{tabular}
\caption{\label{tab:jockey}Top ten jockeys in the random effect from the Forward model, compared to jockey rankings provided by Saratoga, Belmont Park, and Aqueduct in 2022 based on total earnings. NR = not ranked; DNC = did not compete in 2022.}
\end{table}
\end{center}

\subsection{Horse Profiles}\label{sc: profile}

In this section we discuss the estimated forward speed profiles. One of the more complicated choices in building a model like this is to determine the number of knots for the spline as well as the enforced smoothness. In general, we would expect underlying speed profiles to be reasonably smooth and we want to be careful to not overfit the data especially since we expect other effects to explain significant portions of the data variance. \\

\begin{figure}
\begin{subfigure}[a]{0.45\textwidth}
  \includegraphics[scale = .17]{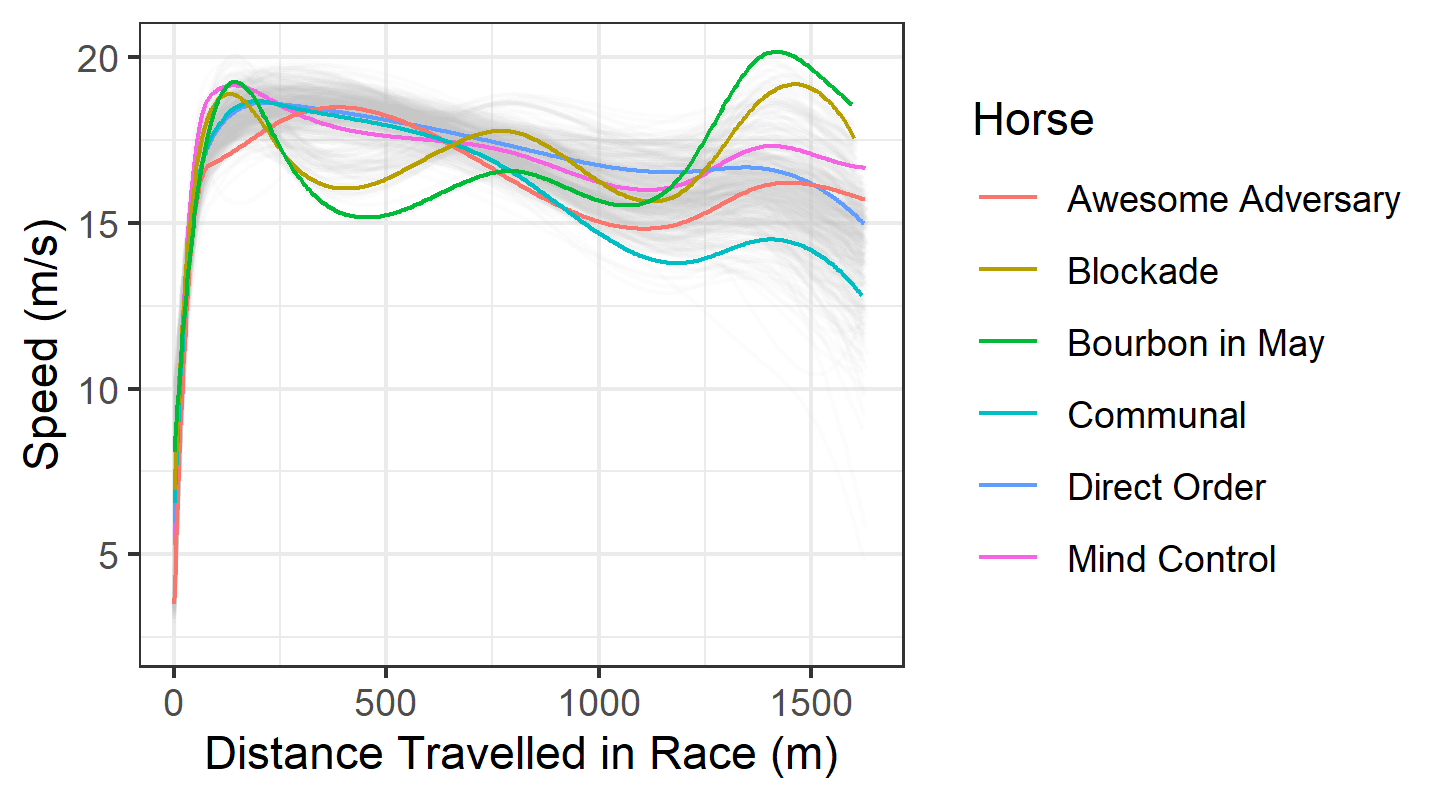}  
  \caption{Forward speed profiles with six selected horses highlighted.}\label{fig:ind horse}
\end{subfigure}
\hfill
\begin{subfigure}[a]{0.45\textwidth}
  \includegraphics[scale = .17]{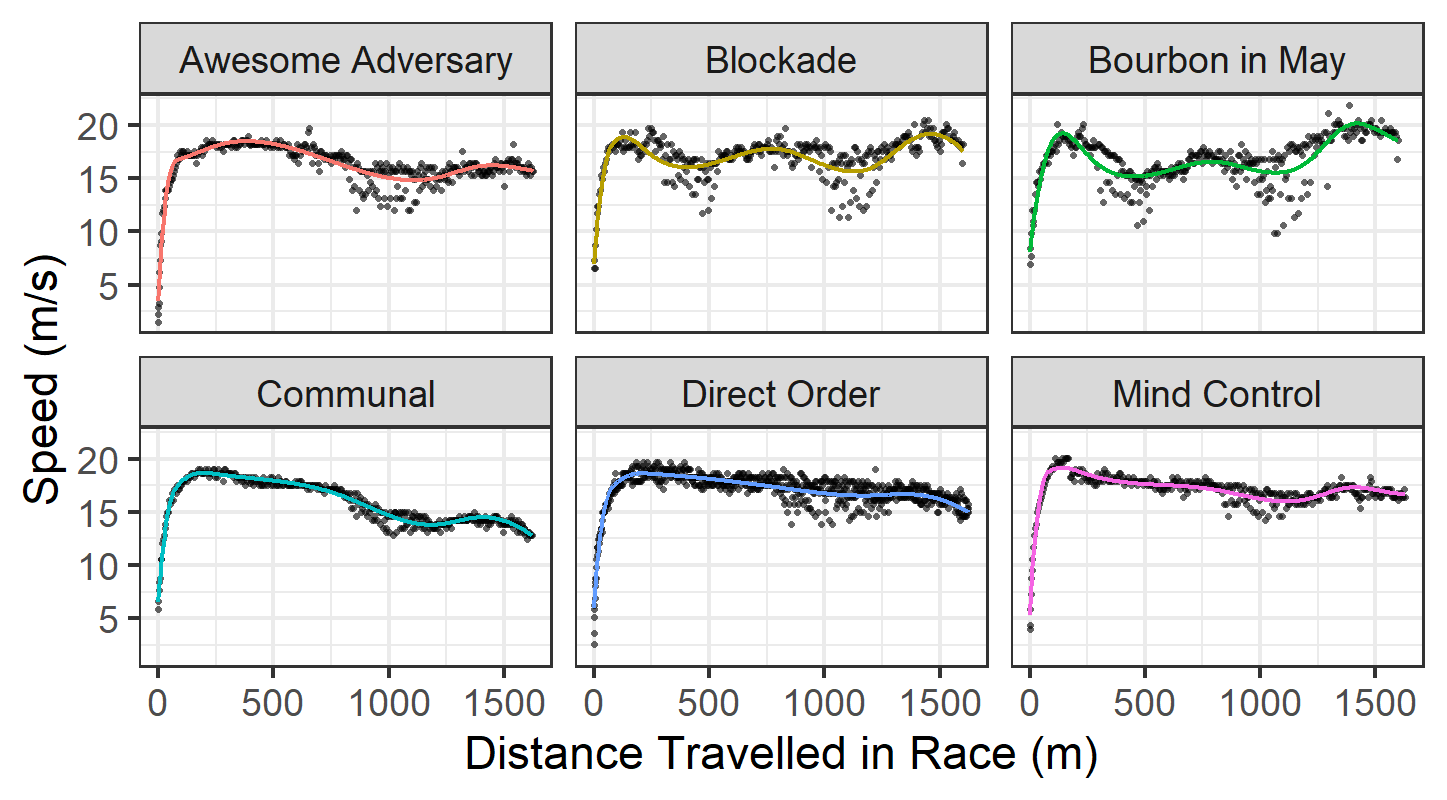} 
  \caption{Six selected horse profiles compared to the observed data from all races.}\label{fig:ind data}
\end{subfigure}
\caption{Horse profile plots from simplified model used to determine the spline complexity and knot locations.}
\end{figure}

As described in Section \ref{sc:mod and sim}, we fit a hierarchical cubic b-spline to incorporate a horse effect into the forward movement model. There are still a number of additional choices required to fit such a spline model including the number of knots and the placement of those knots. Spline knots can be categorized into two types -boundary and internal. The boundary knots anchor the start and end of the smooth curve. The internal knots divide the cumulative distance (i.e. distance along the track) between the boundary knots into segments and influence the shape of the b-spline curve within each segment. They allow us to better capture differences in a horse’s speed tendencies for different distances into a race. Since we consider only 1-mile races, we choose 0 and 1650m as boundary knots. Internal knot placement had to be chosen to accommodate all horses and tracks, as the knots were kept constant across all horses and each horse ran on multiple tracks. To select these hyper-parameters, we used a combination of visual assessment and a leave-one-out cross-validation (LOOCV) approximation using the loo package in Rstan \cite{loo} on a subset of the horse data using a simplified model. Plots of spline fit for that subset of horse were used to choose the degree of the b-spline (3) and the number of internal knots required to capture trends in speed (5), and to obtain reasonable candidate sets of internal knots. LOOCV was used to determine the best choice among these candidate sets. We chose internal knots of 90, 250, 800, 1207, and 1375m. The estimated horse profiles from the simplified model can be seen in Figure \ref{fig:ind horse} and a selection of profiles compared to the observed data can be seen in Figure \ref{fig:ind data}.\\

\begin{figure}[!htb]
\centering
\includegraphics[width=0.5\textwidth]{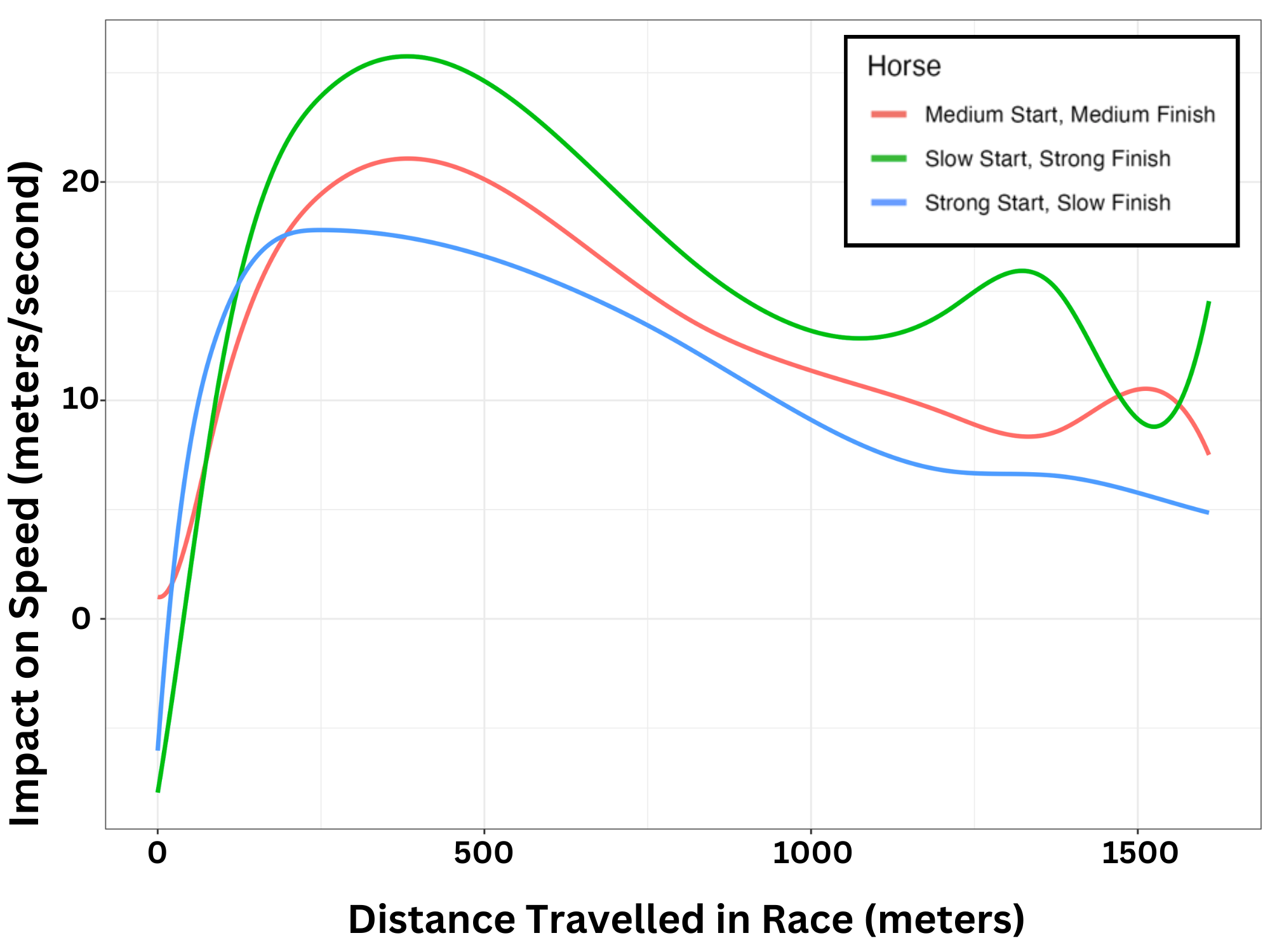}
\caption{\label{fig:horseProfs} The three identified horse profiles using Ward's Linkage. The blue cluster represents horses that are able to reach their top speed the quickest but fail to maintain that speed over the course of the race. The greeen cluster represents horses that are slow out of the gate but reach the highest top speed and finish strong. The red cluster represents horses in the middle of the pack who both start the at middling acceleration and finish at a middling pace.}
\end{figure}

There are numerous inferences and further analyses that can be made with respect to the estimated latent parameters. One particular example we explore here is clustering latent effects on the final fitted model to understand various horse speed profiles in our dataset. This can help to analyze horse tendencies and strengths with respect to the distance travelled along the track. We performed hierarchical clustering with Ward’s linkage on all horses that competed in at least 5 races in our data. As a result, we obtain 3 clusters which we label “Strong Build, Slow Finish” (blue), “Medium Build, Medium Finish” (red), and “Slow Build, Strong Finish” (green) as shown in Figure~\ref{fig:horseProfs}. The Strong Build, Slow Finish group has exceptional acceleration over the first 100m but slowly trails off throughout the race. The Medium Build, Medium Finish group takes a bit longer to reach its top speed but maintains it well throughout the race. The Slow Start, Strong Finish group takes longer to reach the top speed but holds a higher impact on speed throughout most of the race and has an additional burst of energy at the end of the race.

Additionally, we provide the resulting dendrogram from our hierarchical clustering results in the Appendix in Figure~\ref{fig:horseClusts}. This gives us a sense of the relationship between horses with respect to their speed patterns. We logically find a large proportion of the horses analyzed belong to the Medium Build, Medium Finish group as these represent the horses that possess a consistent race pace.

\subsection{Counter-factual Simulations}

Fully generative models for multi-competitor races open up new possibilities for analysis. In particular one can simulate races or strategies that are not observed. As a simple example of the kinds of insights that counter-factual simulations can provide we show how one could estimate starting lane effects. It is important to control for competitor strength when attempting to estimate lane effects. In this example we randomly select six horse/jockey pairs and simulate races between them in all possible lane assignments. With six horse/jockey pairs, there are $6! = 720$ different lane combinations and for each lane combination we simulate 100 races from the posterior predictive distribution.

\begin{figure}[H]
 \centering
    \includegraphics[scale = .2]{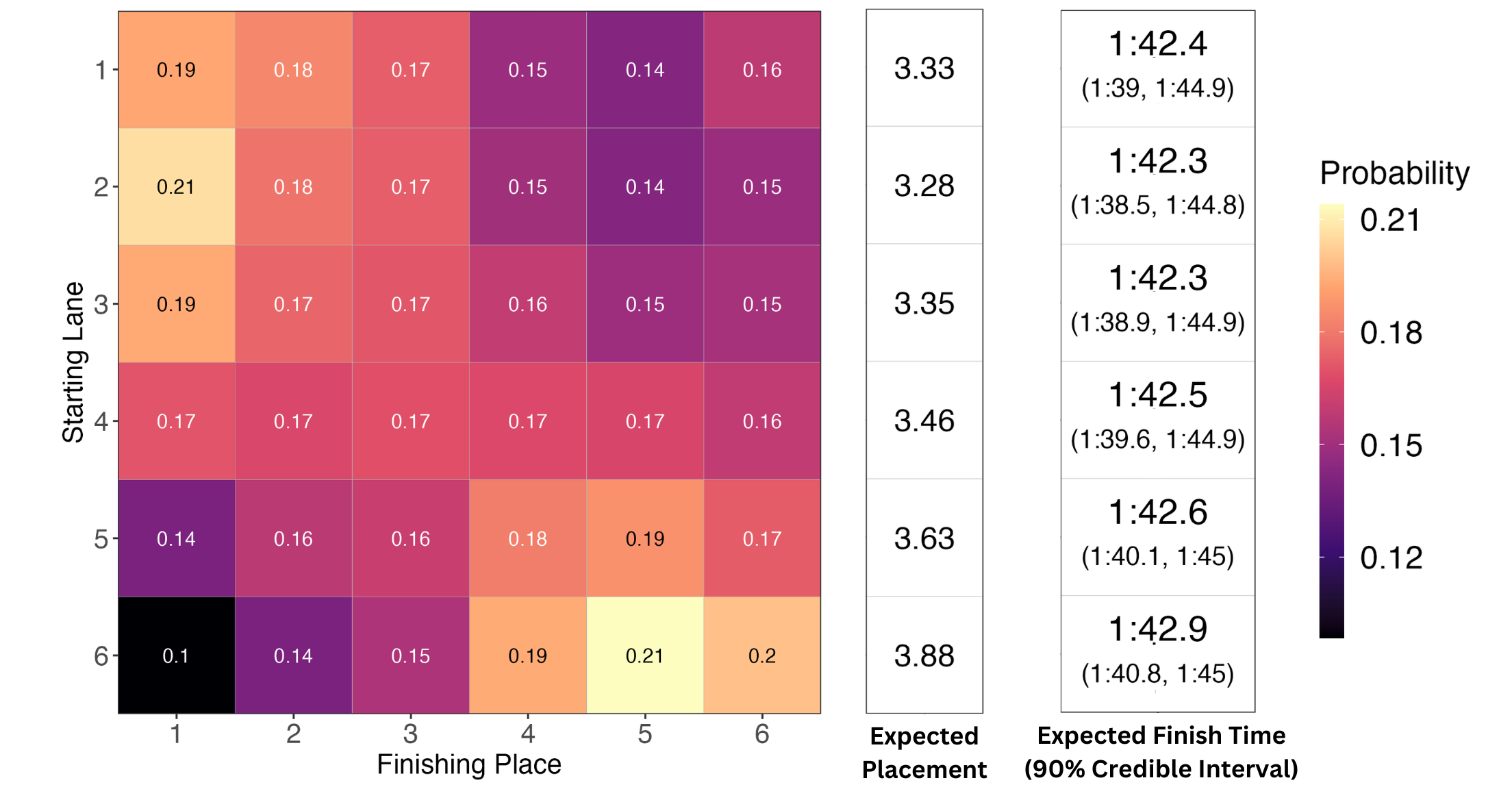}
    \caption{Posterior predictive expected ranks, finishing probabilities, and finishing times from $720\times 100$ posterior predictive simulations from the starting lane experiment. Coloured square heat-map on the left represents placement probabilities for all combinations of finishing placement x-axis and starting lane y-axis. The white column in the middle represents the mean finishing placement for each starting lane based on the 72,000 simulations. The white column on the right represents the mean finishing time for each horse along with a 95\% credible interval in brackets based on the 72,000 simulations.}
    \label{fig:cf sim}
\end{figure}

In Figure \ref{fig:cf sim} we see the results of our counter-factual simulation. The second lane has the lowest (best) expected finishing rank at 3.28 as well as the highest probability of finishing in first, 0.21. The two lanes adjacent to the second lane, the first and third, have the next lowest expected ranks and finishing probabilities. From the fourth lane to the sixth lane the expected rank increases and the probability of placing first, second, and third decreases monotonically. The sixth lane seems to present the largest disadvantage and risk with the highest expected rank of 3.88. This is largely driven by the sixth lane having an elevated chance of finishing last and significantly decreased probability of finishing first. Notice, however, the probability of finishing second through fourth is similar to nearby lanes. Another advantage of fully generative models is that analysis can go beyond simply estimating expectations as we have presented here. To better understand why the sixth lane is disadvantageous, for example, one could study the properties of the simulation draws themselves to identify patterns and characteristics leading to low results. Those patterns could be further stratified with respect to different racing styles or horse characteristics since we might expect effect heterogeneity with respect to lane effects.\\

Overall, we see that even in this simple setting, with a relatively straightforward simulation set-up and question, that fully generative simulations can reveal and help us better understand non-obvious complexities. This is really valuable to competitors, race organizers, and other stakeholders especially in competitions where strategy plays a large role such as horse racing.\\

Even with respect to estimating lane effects, there are several ways to modify the simulation experiment in accord with specific inferential or predictive goals. For example, one might argue that this simulation estimates a particular conditional average treatment effect (CATE) which is specific to the particular horse and jockey pairs that we chose. Depending on the question at hand, it may be more appropriate to estimate a marginal average treatment effect (ATE) by averaging over many horse and jockey pairs racing against themselves. The ATE, for example, might better answer the question about lane effects from the perspective of a race organizer wanting to either keep races fair or to appropriately reward competitors having done well in previous rounds or competitions. On the other hand, by choosing the competitors carefully according to known abilities or tendencies, one could estimate a (potentially) more informative CATE for developing and understanding optimal strategy with respect to a specific competitor or type of competitor.

\section{Conclusion and Future Work}

In the work we propose a generative model compatible with multi-competitor races with available frame-level tracking data. We show how this class of models can capture some of the important within-race dynamics of such races including tactical movements and strategies. The key to these models is modelling total movement as a function of perpendicular and lateral movement at each time step. We demonstrate how this can be applied to the context of one-mile horse races using high-resolution tracking data provided by the NYRA and NYTHA.\\

The contributions of this paper are three-fold. First, we estimate within-race competitor-specific coefficients which vary smoothly over the course of the race and separate these effects from both observed race-level coefficients such as jockey and track effects as well as intra-race factors such as drafting and other dynamic effects. Second, we show how these models can be used to generate computationally feasible posterior predictive simulations of entire races for any starting positions and competitors for which we have suitable data. These simulations can be used to generate instantaneous notions of value analogous to those available through the EPV framework in continuous sports like basketball and soccer. Measures of continuous value can then be further analyzed to study tactics or to attribute value to competitor decisions for example. Finally, the generative nature of the models allows one to simulate counter-factual scenarios to understand probable results given alternative strategies not observed in actual competitions or to study race effects such as our lane effect case-study. This can be especially powerful in collaboration with domain experts who are able to adequately describe potential strategies or research questions of interest.\\

The proposed class of generative models is sufficiently general to apply to any multi-competitor race where there is available tracking data and competitor motion is adequately represented by forward and lateral movements such as most track based events. This is not to say that adaptation to other race settings can be done without care. One of the desirable features of many horse races is the combination a high sampling rate of the data ($\sim 4hz$) and the relatively short nature of many events (e.g the canonical one-mile event takes on the order of 90 seconds and 350-400 frames). High sampling rates allow one to accurately capture instantaneous features such as drafting and relatively short overall races make simulations computationally feasible to generate. Some multi-competitor races will not have access to this level of data or may be significantly longer that simulations on this scale of granularity may not be possible. For example, a Tour de France stage is often on the order of several hundred kilometers and may take many hours to complete. Such cases will require some adaptations to be feasible.\\

There are two broad classes of solutions to these problems - sampling schemes and emulators which may be used in combination with each other. The most natural solution to dealing with races which are an order of magnitude larger in terms of number of frames (or number of competitors or both) is to coarsen the sampling until it is a manageable size. In some races it may be necessary to make the coarsening quite significant. The trade-off with coarsening is that features like drafting cease to be instantaneous measures and additionally this lack of granularity may show up in some measures of uncertainty and impact the granularity of the questions that can be answered by the simulation. In some cases, some of the drawbacks may be partially overcome by coarsening over meaningful subsections of the race. In road cycling, for example, stages are often classified into particular subsections of flats, hills, and descents. It may be sufficient to capture things like rider ability and average drag over these subsections (or further divisions thereof) to generate meaningful inferences, predictions, and simulations. The subsections need not be overly coarse, however. As we demonstrated in our canonical example we were able to simulate feasibly on the order of 400 frames and in many cases dividing a race into several hundred subsections may not be overly restrictive. Another sampling approach can involve down-sampling and interpolating over these subsections rather than modelling the subsections as a distinct unit. Coarsening and down-sampling approaches may also be combined when appropriate.\\

The problem of needing to lower the computational burden of inferences and predictions based on computationally intensive simulations is hardly unique to sports and the tracking data context. Emulators, that is models which approximate the outputs of complex simulations are, in fact, common in physical and social sciences such as physics \citep{kataoka2023machine}, biology \citep{stolfi2021emulating}, and fields like economics or sociology where agent based modelling is employed \citep{angione2020using}. In practice, models like Gaussian processes and neural networks are trained based on outcomes which are some function of the simulation results. In some cases Bayesian methods are preferred when one is interested in capturing the underlying uncertainty in the simulations rather than simply point estimates of expectations \citep{vernon2022bayesian}, however there is an emerging literature aiming to quantify the epistemic and model uncertainty of deeper machine learning methods in the context of emulators \citep{thiagarajan2020designing}. The emulator still requires some number of full simulations to be conducted to develop a sufficient training set for the task at hand to ensure the key features of the richer simulation framework is learned effectively. It is also important to note that several different emulation models may need to be developed to model different outputs for the simulation. For example, returning to the canonical horse racing example, one may need to build a different emulator for predicting race finishing times than if one is investigating the importance of lane assignment on lateral movement in the early stages of the race.

\bibliographystyle{plain}

%\bibliography{Sources}

\section{Appendix}

\subsection{Covariates Table}

\begin{center}
\begin{table}[H]
\centering
\begin{tabular}{@{}ccc@{}}
\toprule
Feature                                       & Type & Description                                                                                    \\ \midrule
\multicolumn{1}{c|}{n\_horses\_inside}        & DWR  & Number of horses to the inside (left).                                                         \\
\multicolumn{1}{c|}{n\_horses\_outside}       & DWR  & Number of horses to the outside (right).                                                       \\
\multicolumn{1}{c|}{n\_horses\_forward}       & DWR  & Number of horses in front.                                                                     \\
\multicolumn{1}{c|}{n\_horses\_backward}      & DWR  & Number of horses behind.                                                                       \\
\multicolumn{1}{c|}{nearest\_inside}          & DWR  & Nearest horse on the inside, in terms of lateral distance.                                     \\
\multicolumn{1}{c|}{nearest\_outside}         & DWR  & Nearest horse to the outside, in terms of lateral distance.                                    \\
\multicolumn{1}{c|}{nearest\_inside\_euclid}  & DWR  & Nearest horse on the inside, in terms of Euclidean distance.                                   \\
\multicolumn{1}{c|}{nearest\_outside\_euclid} & DWR  & Nearest horse to the outside, in terms of Euclidean distance.                                  \\
\multicolumn{1}{c|}{nearest\_forward}         & DWR  & Nearest horse in front, in terms of forward distance.                                          \\
\multicolumn{1}{c|}{prev\_lat\_movement}      & DWR  & Lateral distance travelled in previous frame (LMO)                                             \\
\multicolumn{1}{c|}{is\_drafting}             & DWR  & An indicator for whether the horse is drafting in the current frame.                           \\
\multicolumn{1}{c|}{prop\_energy\_saved}      & DWR  & Total proportion of energy saved due to drafting.                                              \\
\multicolumn{1}{c|}{is\_turn}                 & DWR  & An indicator for whether the horse is going around a turn.                                     \\
\multicolumn{1}{c|}{is\_home\_stretch}        & DWR  & Horse is in the home stretch of the race (LMO)                                                 \\
\multicolumn{1}{c|}{turn\_to\_home\_stretch}  & DWR  & Horse is in the first 10m of the home stretch coming out of turn (LMO)                         \\
\multicolumn{1}{c|}{race\_context}            & RE   & A combination of track type (dirt, turf) and surface condition (fast, good, sloppy, or muddy). \\
\multicolumn{1}{c|}{jockey}                   & RE   & A simple random effect for the jockey.                                                         \\
\multicolumn{1}{c|}{horse\_spline}            & RE   & A hierarchical B-spline describing the movement pattern of each horse (FMO).                   \\ \bottomrule
\end{tabular}
\caption{\label{tab:features}All covariates and effects used in proposed forward and lateral movement models for each combination of horse and jockey during active races. DWR: Dynamic Within-Race Feature. RE: Random Effect. LMO: Lateral Model Only. FMO: Forward Model Only.}
\end{table}
\end{center}

\subsection{Clustering Dendrogram}

\begin{figure}[!htb]
\centering
\includegraphics[width=0.65\textwidth]{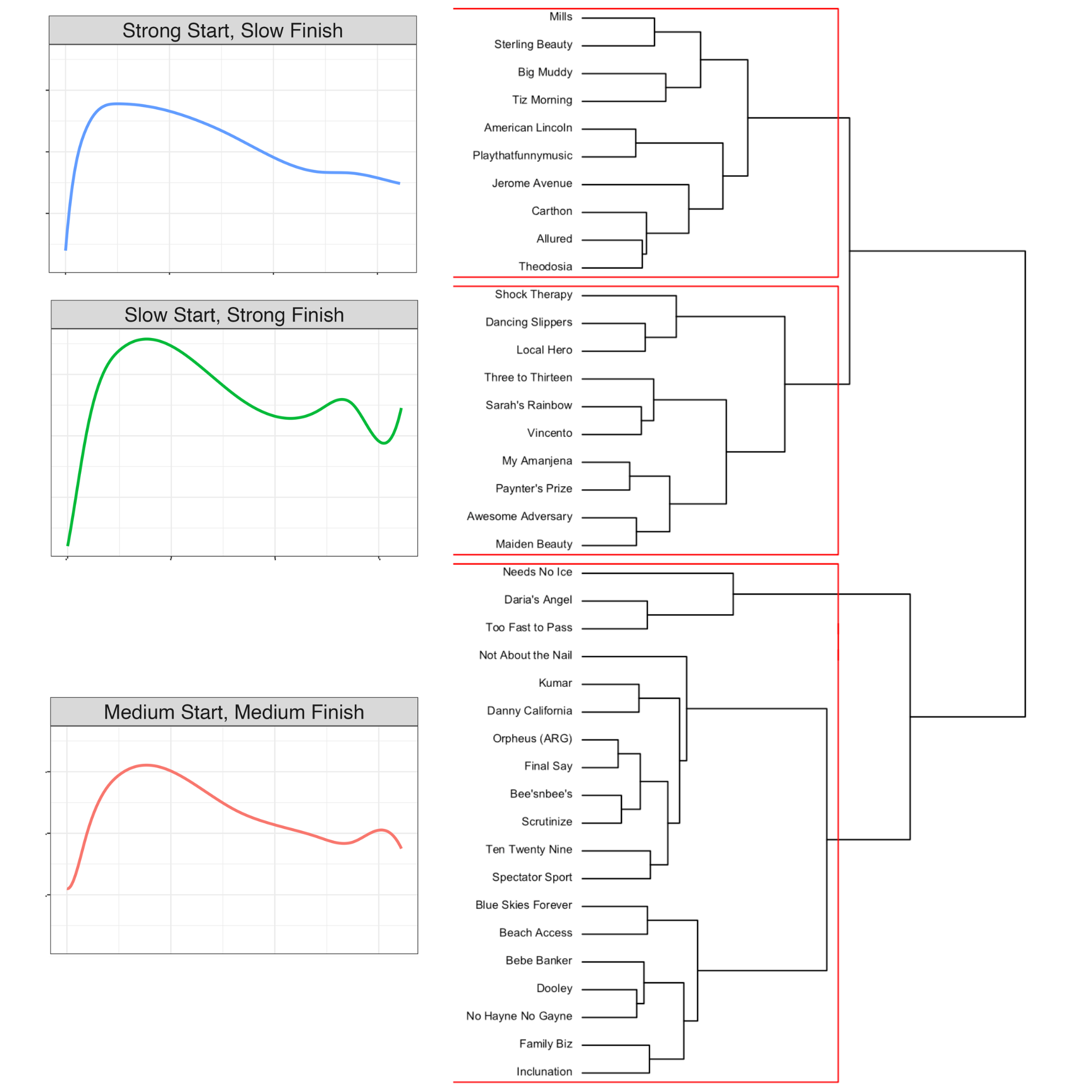}
\caption{\label{fig:horseClusts} A hierarchical clustering dendrogram based on all horses with 5 or more 1 mile races. Red borders are used to divide clusters.}
\end{figure}

\end{document}